\documentclass[sigconf]{acmart}
\acmConference[ICSE 2022]{The 44th International Conference on Software Engineering}{May 21–29, 2022}{Pittsburgh, PA, USA}

\pdfoutput=1
\usepackage{amsmath,amssymb,amsfonts}
\usepackage{textcomp}
\usepackage{xcolor}
\usepackage{makecell}
\usepackage{framed}
\usepackage{algpseudocode}
\usepackage{mathrsfs}
\usepackage{bm}
\usepackage{threeparttable}
\usepackage{array}
\usepackage{graphicx}
\usepackage{subfigure}
\usepackage{balance}
\usepackage{multirow}  
\usepackage[linesnumbered,lined,ruled,commentsnumbered]{algorithm2e}

\def\BibTeX{{\rm B\kern-.05em{\sc i\kern-.025em b}\kern-.08em
    T\kern-.1667em\lower.7ex\hbox{E}\kern-.125emX}}
    
\copyrightyear{2022}
\acmYear{2022}
\setcopyright{acmcopyright}\acmConference[ICSE '22]{44th International Conference on Software Engineering}{May 21--29, 2022}{Pittsburgh, PA, USA}
\acmBooktitle{44th International Conference on Software Engineering (ICSE '22), May 21--29, 2022, Pittsburgh, PA, USA}
\acmPrice{15.00}
\acmDOI{10.1145/3510003.3510213}
\acmISBN{978-1-4503-9221-1/22/05}
    
\begin{document}

\title{Modeling Review History for Reviewer Recommendation:\\A Hypergraph Approach
}


\author{Guoping Rong, Yifan Zhang, Lanxin Yang, Fuli Zhang, Hongyu Kuang, He Zhang}
\affiliation{%
  \institution{State Key Laboratory of Novel Software Technology, Software Institute, Nanjing University}
  \city{Nanjing}
  \state{Jiangsu}
  \country{China}
}
\email{ronggp@nju.edu.cn, yifanzhang590@gmail.com, yang931001@outlook.com}
\email{mg1932016@smail.nju.edu.cn, khy@nju.edu.cn, hezhang@nju.edu.cn}

\begin{abstract}
Modern code review is a critical and indispensable practice in a pull-request development paradigm that prevails in Open Source Software (OSS) development. Finding a suitable reviewer in projects with massive participants thus becomes an increasingly challenging task. Many reviewer recommendation approaches (recommenders) have been developed to support this task which apply a similar strategy, i.e. modeling the review history first then followed by predicting/recommending a reviewer based on the model. Apparently, the better the model reflects the reality in review history, the higher recommender's performance we may expect. However, one typical scenario in a pull-request development paradigm, i.e. one \emph{Pull-Request (PR)} (such as a revision or addition submitted by a contributor) may have multiple reviewers and they may impact each other through publicly posted comments, has not been modeled well in existing recommenders.
We adopted the hypergraph technique to model this high-order relationship (i.e. one~\emph{PR} with multiple reviewers herein) and developed a new recommender, namely~\emph{HGRec}, which is evaluated by 12 OSS projects with more than 87K \emph{PR}s, 680K comments in terms of~\emph{accuracy} and~\emph{recommendation distribution}. The results indicate that~\emph{HGRec} outperforms the state-of-the-art recommenders on recommendation accuracy. Besides, among the top three accurate recommenders,~\emph{HGRec} is more likely to recommend a diversity of reviewers, which can help to relieve the core reviewers' workload congestion issue. Moreover, since~\emph{HGRec} is based on hypergraph, which is a natural and interpretable representation 
to model review history, it is easy to accommodate more types of entities and realistic relationships in modern code review scenarios. 
As the first attempt, this study reveals the potentials of hypergraph on advancing the pragmatic solutions for code reviewer recommendation.
\end{abstract}



\maketitle

\section{Introduction}
As a popular software practice, code review is believed to be paramount to software quality for both commercial projects and Open Source Software (OSS) projects~\cite{bosu2016process, thompson2017large, shimagaki2016study, sadowski2018modern}. Through manually scrutinizing source code, reviewers aim to identify possible issues or improvement opportunities and thereby prevent issue-prone code snippets from being incorporated into project repositories~\cite{beller2014modern}. In addition to secure quality, code review is also helpful in knowledge dissemination, team collaboration~\cite{bacchelli2013expectations, bosu2013impact, tsay2014let, macleod2017code}, etc. However, studies show that code review highly relies on the experience and knowledge of reviewers~\cite{kononenko2016code, mcintosh2014impact}, which implies that the identification of suitable reviewers is crucial to review efficacy.

Nowadays, an informal, asynchronous and tool-based code review practice that is known as Modern Code Review (MCR) is widely adopted in software development~\cite{mcintosh2016empirical}. In the OSS community, MCR is an essential step in a so-called pull-request development paradigm~\cite{yu2014should, yu2015wait}, where developers make changes to some code snippets and submit a \emph{Pull-Request (PR)} to the project repository. Then potential reviewers (including project owners) examine the~\emph{PR} and provide feedback through an issue tracking system (e.g., JIRA\footnote{https://www.atlassian.com/software/jira}, Gerrit\footnote{https://www.gerritcodereview.com/}, etc.); if the issues related to the \emph{PR} were properly addressed, one project owner then merges the~\emph{PR} into the project repository~\cite{kononenko2018studying, zampetti2019study}. Studies indicate that MCR quality is subject to many factors, among which the reviewers' expertise and workload can make a significant difference~\cite{baysal2013influence, ruangwan2019impact, kononenko2016code}.
In this sense, it is also crucial to find suitable reviewers for a certain ~\emph{PR}, especially in the context of the OSS development where the potentially massive participants are usually geographically distributed and not necessarily known to each other.
In fact, as Thongtanunam et al. pointed out, inappropriate assignment of code reviewer may take 12 days longer to approve a code change in OSS development, thus a recommendation tool is necessary to speed up a code review process~\cite{thongtanunam2015should}.

In the past decade, researchers have worked out a number of reviewer recommendation approaches (recommenders) in order to automatically assign a ~\emph{PR} to potentially suitable reviewers. In general, recommenders are developed by following a similar strategy, i.e., to predict/recommend a reviewer based on a model from historical reviews. For example, heuristic-based recommenders model the review history by mining simple rules based on the relationships 
among source files, revisions, and participants. That is, whoever most frequently revised or reviewed the source code snippets included in a certain~\emph{PR} previously should be recommended to perform the review first. Learning-based recommenders model the review history by machine learning techniques~\cite{jeong2009improving, jiang2015coredevrec} and then use the trained models to determine the most potentially suitable reviewers.

It is widely agreed that models play a vital role in all recommenders. However, the models behind heuristics-based recommenders are combinations of simple rules reflecting relationships, which is very likely to miss crucial information (e.g., the mutual impacts among reviewers). This may be one of the reasons that most heuristics-based recommenders can not achieve satisfactory recommendation accuracy~\cite{lipcak2018large, jiang2015coredevrec}. Meanwhile, the learning-based recommenders may be able to process more information of the review history yet the low interpretable models behind and heavy workload on feature engineering also prevent them from evolving to quickly adapt themselves to new situations.
Recently, some researchers began to apply graph techniques to model the relationships among entities such as source code, participants, \emph{PR}s, etc. As the modeling data structure, a graph is able to support more sophisticated heuristics algorithms. Besides, it is also able to support multiple machine learning algorithms and improve model interpretability~\cite{yu2016reviewer, ying2016earec}. 

However, since a single edge in an ordinary graph can only associate two vertexes, it is difficult (if not impossible) to model a common phenomenon in OSS development, i.e., one \emph{PR} may involve multiple reviewers. As a result, most graph-based recommenders also have not presented satisfactory recommendation accuracy~\cite{jiang2017should, xia2017hybrid}. 

\begin{figure}[!htbp]
  \centering
  \includegraphics[width=0.9\linewidth]{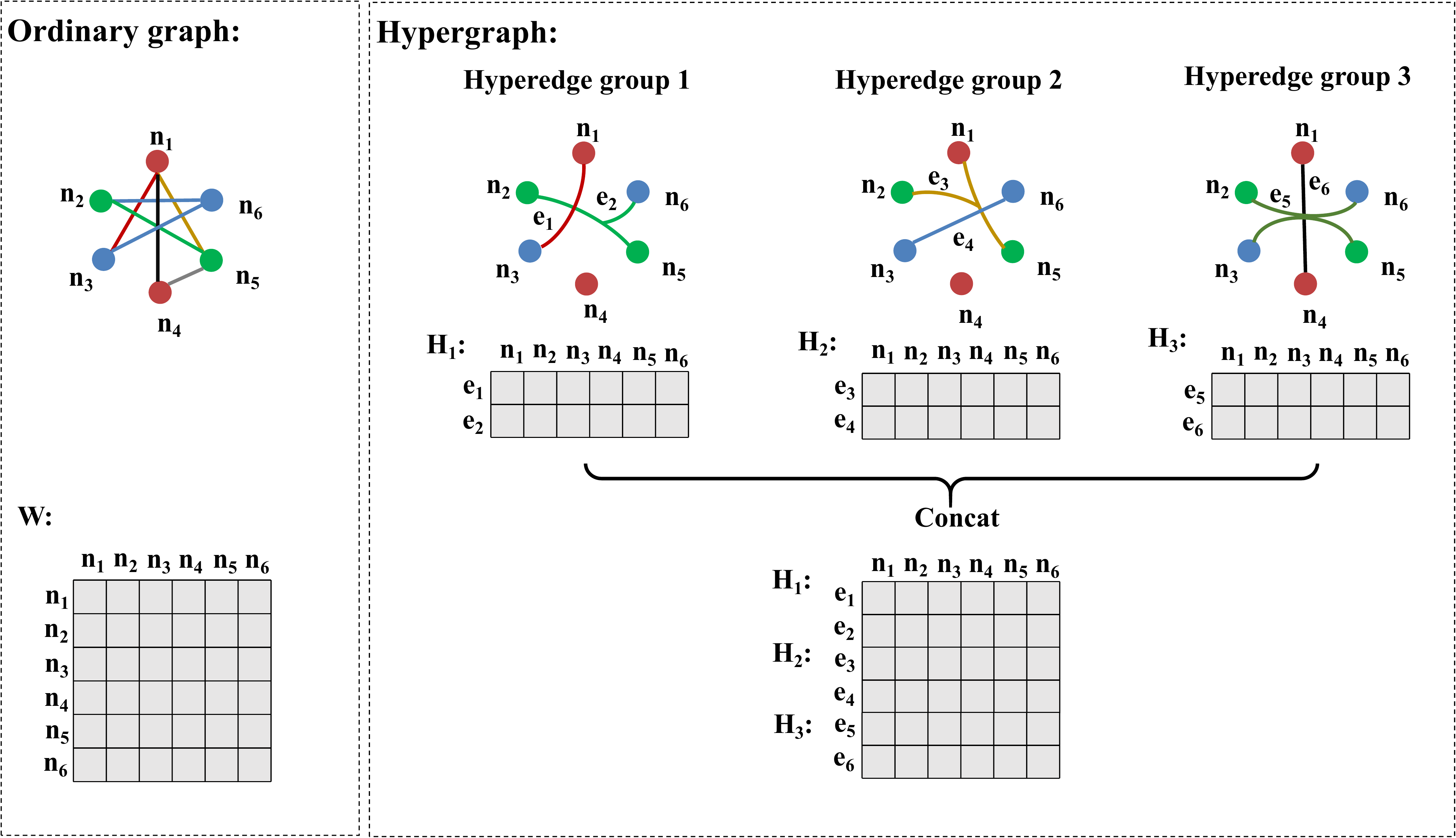}
  \caption{Ordinary graph and hypergraph}
  \label{FIG:compGAndHG}
\end{figure}

\begin{figure}[!htbp]
  \centering
  \includegraphics[width=0.9\linewidth]{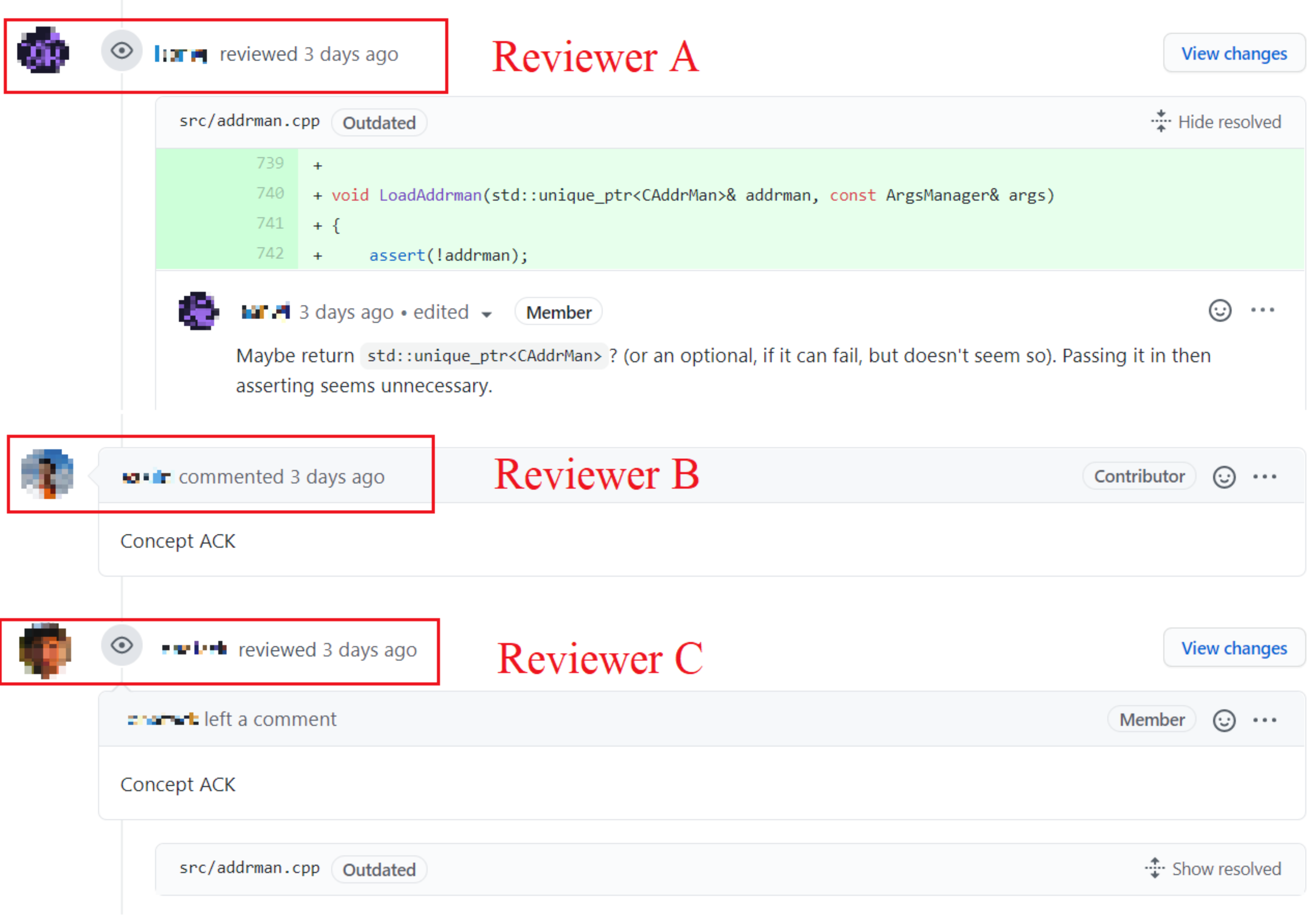}
  \caption{One \emph{PR} involves multiple reviewers in practice}
  \label{FIG:reviewWithReviewers}
\end{figure}

 Recently, a technique namely hypergraph has been utilized to model the complex relationships among multiple entities. Briefly, a hypergraph is a generalization of an ordinary graph in which an edge can associate any number of vertexes. For example, as shown on the right of Figure~\ref{FIG:compGAndHG}, a hyperedge $e_2$ simultaneously connects three vertexes ($n_2$, $n_5$, $n_6$). In contrast, in an ordinary graph, an edge connects exactly two vertexes (shown on the left of Figure~\ref{FIG:compGAndHG}). 
Take Figure~\ref{FIG:reviewWithReviewers} as a review example, there are three reviewers, namely \emph{Reviewer A}, \emph{Reviewer B} and \emph{Reviewer C} having reviewed the identical \emph{PR}. In an ordinary graph, the review history is modeled as reviewer pairs, i.e., \emph{Reviewer (A, B)}, \emph{Reviewer (B, C)} and \emph{Reviewer (A, C)} have reviewed one same \emph{PR}. However, the fact that \emph{ Reviewer A, B} and \emph{C} actually have reviewed the same \emph{PR} is not able to be reflected in an ordinary graph. Since certain form of familiarity (e.g., similar review experience/history) forms the basis for most reviewer recommenders, the loss of this information will inevitably impact the recommendation performance.
In contrast, since multiple vertexes can be included in one edge, hypergraph offers more natural approaches to model the review history portrayed in Figure~\ref{FIG:reviewWithReviewers}, which provides more information for recommenders to perform recommendation. 



In this study, we applied the hypergraph technique to model the aforementioned complex relationship among various entities in a natural and interpretable way. Based on this model, we also developed a new reviewer recommender, namely~\emph{HGRec} to explore the feasibility and effectiveness of this strategy. An extensive empirical study based on 12 OSS projects with more than 87K \emph{PR}s and 680K review comments indicates the superiority of \emph{HGRec} in terms of recommendation accuracy as well as workload balance among reviewers. The contributions of this study can be highlighted as below.
\begin{itemize}
    \item To the best of our knowledge, this is the first effort that hypergraph is used to model code reviews as well as complex relationships among participants, e.g., one reviewer may be affected by others' comments.
    \item We developed a new recommender, i.e.,~\emph{HGRec}, based on hypergraph technologies.
    \item We empirically evaluated~\emph{HGRec}, the results indicate that~\emph{HGRec} not only outperforms the state-of-the-art recommenders in terms of recommendation accuracy, but to some extent mitigates the workload congestion issue.
   
\end{itemize}

\section{Related Work}
\subsection{Code Reviewer Recommendation}
\subsubsection{Recommenders}
Automated reviewer recommendation has attracted a lot of attention in the past decade. A number of recommenders have been proposed, which follow a similar strategy in general, i.e., modeling review history and use the result model to recommend a new reviewer. In general, there are three main types of recommenders according to different modeling approaches, i.e. the heuristics-based, learning-based, and graph-based recommenders, respectively.

\paragraph[short]{Heuristics-based recommenders}
This type of recommenders suggests new reviewers with simple heuristic rules. For example, Thongtanunam et al.~\cite{thongtanunam2014improving} proposed a recommender based on file path similarity, which subsequently evolved into \emph{RevFinder}~\cite{thongtanunam2015should}.
The~\emph{RevFinder} is based on the similarity between the file paths of a previous \emph{PR} and a new \emph{PR}. Zanjani et al.~\cite{zanjani2015automatically} developed a recommender (\emph{cHRev}) that determines candidates on a basic premise that the reviewers who have reviewed target code snippets before are most likely to be recommended. Rahman et al.~\cite{rahman2016correct} proposed a recommender (\mbox{\it CORRECT}) that utilizes external library similarity and technology expertise similarity of reviewers, which provides a possibility for cross-project reviewer recommendation. Jiang et al.~\cite{jiang2017should} analyzed several attributes related to the code review and found that activeness-based recommender (\emph{AC}) performed the best. Other rules adopted in the  heuristics-based recommenders include~\emph{Line 10 Rule}~\cite{schuler2008mining},~\emph{Expertise Recommender}~\cite{mcdonald2000expertise},~\emph{Code Ownership}~\cite{girba2005developers} and~\emph{Expertise Cloud}~\cite{alonso2008expertise}, etc. Usually, the \emph{`models'} used by the  heuristics-based recommenders are merely simple statistics or comparison results on the original review history. Most heuristic-based recommenders are easy to understand. However, research indicates that most of them suffer from low accuracy. Moreover, it is usually hard to add more elements (information) to enhance the models based on simple heuristic rules, which impacts their evolvability.

\vspace{-0.1cm} 
\paragraph[short]{Learning-based recommenders}
This type of recommenders assumes that the~\emph{PR} profile and reviewers' personal expertise can be automatically learned from the review history by training. Among them, Support Vector Machine (\mbox{\it SVM}), Random Forest (\emph{RF}), and Bayesian Network (\emph{BN}) are widely applied~\cite{jeong2009improving, jiang2015coredevrec, hannebauer2016automatically, jiang2019should}. de Lima J{\'u}nior et al.~\cite{de2015developers} investigated several kinds of learning-based recommenders, including Na\"ive Bayes (\emph{NB}), Decision Tree (\emph{J48}),~\emph{RF}, and Sequential Minimal Optimization (\mbox{\it SMO}) and found that~\emph{RF} outperforms others in terms of recommendation accuracy. In general, learning-based recommenders usually perform better than simple heuristics-based recommenders, however, the models behind these recommenders need a heavy workload on feature engineering, training, and long-term maintenance. Besides, they are normally not interpretable also, which becomes a barrier for future extension and improvement for the recommenders.

\vspace{-0.1cm} 
\paragraph[short]{Graph-based recommenders}
Recently, graph techniques have been adopted to model the review history~\cite{sulun2021rstrace, sulun2019reviewer, liao2019core, yu2016reviewer, ouni2016search}, through which personal profiles and social relationships or networks between developers and reviewers are thus formalized into graph vertexes and edges. Using graph as the model, both sophisticated heuristics and learning algorithms can be used to design recommenders. For example, 
Yu et al.~\cite{yu2016reviewer} found that developers who share common interests with a \emph{PR} originator are potentially suitable reviewer candidates. Liao et al.~\cite{liao2019core} combined \emph{PR} topic model with social networks to build the connections between collaborators and \emph{PR}s. S{\"u}l{\"u}n et al.~\cite{sulun2019reviewer} used software artifact traceability graphs to recommend reviewers who potentially are familiar with a given artifact.


\vspace{-0.1cm}

\subsubsection{Recommendation distribution}
\label{Recommendation distribution}
The rationale behind nearly all the recommenders implies that one reviewer who conducted the most reviews in the history tends to be recommended in a future review. As a matter of fact, it is common that a few core reviewers took over the most workloads on code review~\cite{yang2018revrec},
which becomes a severe issue of ``workload congestion'' for some core reviewers, leading to review overload for these core reviewers~\cite{mirsaeedi2020mitigating}. Recent studies have proposed some solutions to alleviate the workload congestion. Asthana et al.~\cite{asthana2019whodo} proposed a recommender (\emph{WhoDo}) where reviewers' scores are reduced by his/her incomplete~\emph{PR}s so as to decrease his/her chance to be recommended.
Al-Zubaidi et al.~\cite{al2020workload} presented a workload-aware recommender (\emph{WLRRec}) by utilizing~\emph{NSGA-II}, a multi-objective search-based approach to address two main objectives -- maximizing the chance of participating in a review and, minimizing the skewness of review workload distribution. Rebai et al.~\cite{rebai2020multi} balanced the conflicting objectives of expertise, availability, and history of collaborations with multi-objective search techniques. Mirsaeedi et al.~\cite{mirsaeedi2020mitigating} systematically take expertise, workload, and knowledge distribution for collaborators in recommending new reviewers.

In short, the workload congestion issue has raised wide concern in the research community on reviewer recommenders and should not be neglected in designing and evaluating recommenders.

\subsection{Hypergraph Approach for Software Engineering}
A hypergraph is an extension of the ordinary graph that consists of multiple vertexes and hyperedges, which can depict the high-order relationships among entities~\cite{duchenne2011tensor, zhang2013feature}. Therefore, unlike the pairwise relationships depicted in an ordinary graph, hypergraph has the ability to express complex relationships in the real world, which prevents information loss as far as possible~\cite{lin2005simplicial, zhao2018learning, ouvrard2018hypergraph, luqman2019study}. This merit enables hypergraph techniques to be used in some software engineering scenarios. For example, G{\"o}de et al. \cite{gode2011frequency} used hypergraph-based models on cloned code fragments and analyzed clone evolution in mature projects. Thom{\'e} et al. \cite{thome2017search} used hypergraph to implement a search-driven string constraint solving algorithm to detect vulnerabilities in the program. Jiang et al. \cite{jiang2019inferring} used hypergraphs to represent code and implemented a framework for interring program transformations. While the studies that use hypergraph techniques to model the complex relationships among software artifacts are not rare, to the best of our knowledge, this technique has never been used in reviewer recommendation, which usually involves both entities such as humans and artifacts as well as the complex and high-order relationships among different entities.
This motivates the hypergraph-based recommender (i.e., \emph{HGRec}) that is proposed in this study.

\begin{figure*}[htp]
\centering
\includegraphics[width=0.85\linewidth]{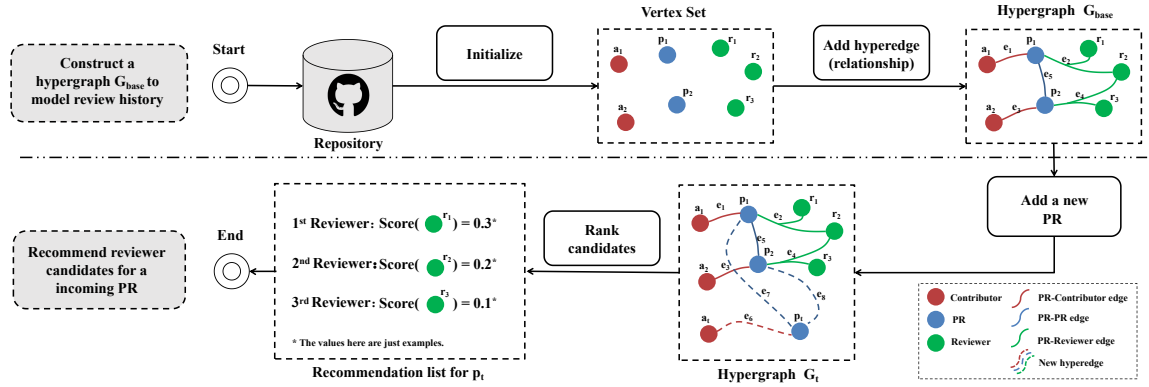}
\caption{Overview of HGRec}
\label{fig:recommender_overview}
\end{figure*}

\section{Approach}
There are two major steps to design and implement \emph{HGRec}, i.e., hypergraph construction and reviewer recommendation, respectively. In this section, we elaborate these two steps in detail.


\subsection{Approach Overview}

Figure~\ref{fig:recommender_overview} depicts the two major steps of \emph{HGRec}. The top segment shows the process to construct a base hypergraph ($G_{base}$), which is based on the review history retrieved from project repositories; the bottom of Figure~\ref{fig:recommender_overview} presents the process to recommend potentially suitable reviewers for an incoming new \emph{PR}, say $p_{t}$. The basic idea is to add $p_{t}$ and corresponding contributor ($a_{t}$) to the existing hypergraph $G_{base}$ to form a new hypergraph $G_{t}$ using the similar strategy to construct $G_{base}$. Then a hypergraph-based search strategy which calculates vertex score using a localized first-order approximation~\cite{bu2010music} is applied to rank and recommend candidate reviewers. Details of the hypergraph construction and reviewer recommendation will be elaborated in the following subsections.

\begin{table}[!htb]
\scriptsize
\caption{Key notations}
    \label{tab:notations}

\begin{threeparttable}
    \begin{tabular}{p{4cm}p{7cm}}
\toprule
\multicolumn{1}{c}{Notations}           & \multicolumn{1}{c}{Descriptions}  \\
\midrule
\multicolumn{1}{l}{$\mbox{\it PRS}$}           &  \multicolumn{1}{l}{the set of  \emph{Pull-Request(PR)}s with $n$ \emph{PR}s at first} \\
\multicolumn{1}{l}{$p_{i}, i \in [1 ... n] $ }           &  \multicolumn{1}{l}{a \emph{PR} in $\mbox{\it PRS}$} \\
\multicolumn{1}{l}{$a_i$}           &  \multicolumn{1}{l}{the contributor of \emph{PR} $p_{i}$} \\
\multicolumn{1}{l}{$R_i$}           &  \multicolumn{1}{l}{the set of reviewers to  \emph{PR} $p_{i}$} \\
\multicolumn{1}{l}{$r_{i_{j}}, j \in [1 \dots m] $}           &  \multicolumn{1}{l}{a reviewer in $R_i$} \\
\multicolumn{1}{l}{$F_i$}           &  \multicolumn{1}{l}{the set of changed file paths involved in \emph{PR} $p_{i}$} \\
\multicolumn{1}{l}{$f_{i_{k}},k \in [1 \dots l]$}           &  \multicolumn{1}{l}{a file path in $F_i$} \\
\multicolumn{1}{l}{$G$}           &  \multicolumn{1}{l}{the hypergraph constructed based on $\mbox{\it PRS}$} \\
\multicolumn{1}{l}{$V$}           &  \multicolumn{1}{l}{the set of vertexes in $G$} \\
\multicolumn{1}{l}{$E$}           &  \multicolumn{1}{l}{the set of hyperegdes in $G$} \\
\bottomrule
\end{tabular}
\end{threeparttable}
\end{table}

In order to eliminate ambiguity, we first define some key notations in Table~\ref{tab:notations}. To be specific, the review history of an OSS project is represented by a set of \emph{PR}s ($\mbox{\it PRS}$), including the contributors ($a_i$), reviewers ($R_i$) and changed file paths ($F_i$)  involved in each \emph{PR} ($p_{i}$). A hypergraph ($G_{base}$) is used to model the review history, based on which, a new hypergraph ($G_{t}$) is generated by adding an incoming new \emph{PR}, say $p_t$, to $G_{base}$.


\subsection{Hypergraph Construction}
Intuitively, for a target \emph{PR}, the adjacent \emph{PR}s in terms of file paths share certain similarities regarding content or function, which may also be able to reflect the similarity regarding experience and familiarity towards the target \emph{PR} among the contributors and reviewers involved in these \emph{PR}s. Using hypergraph the relationships among different entities involved in these \emph{PR}s can be created in a succinct and natural representation. Two major steps are included to construct a hypergraph, i.e. the architecture building and the edge weight, respectively. As shown in 
Algorithm~\ref{algo_graphConstruction}, function~\emph{Construction}  depicts the former step, while the latter step is described by the function~\emph{BuildEdge}.

\begin{algorithm}
    \scriptsize
    \caption{Hypergraph Construction}
    \label{algo_graphConstruction}
    \SetKwFunction{BuildEdge}{BuildEdge}
    \SetKwFunction{AddVertex}{AddVertex}
    \SetKwFunction{SetWeight}{SetWeight}
    \SetKwFunction{CaculateWeight}{CaculateWeight}
    \SetKwFunction{SortWeightAddFilter}{SortWeightAddFilter}
    
    \SetKwInOut{Input}{input}\SetKwInOut{Output}{output}
    \Input{PR Set $\mbox{\it PRS}$ \\  Contributor List  $A = \langle a_{1},a_{2},\dots,a_{n} \rangle$  \\  Reviewer Set $R=\{R_{1},R_{2},\dots,R_{n}\}$}
    \Output{Hypergraph $G_{base}=\{V_{base},E_{base}\}$}
    \BlankLine
    \SetKwFunction{FMain}{Construction}
    \SetKwProg{Fn}{Function}{}{}
    \Fn{\FMain{$\mbox{\it PRS}, A, R$}}{
        $V_{base} \gets \oslash$; $E_{base} \gets \oslash$\;
        \For {$p_{i} \in \mbox{\it PRS}$}{
            $V_{base} \gets \{V_{base}\cup \{p_{i}\} \cup \{a_{i}\} \cup R_{i}\}$  \tcp*[l]{add vertexes}
            Edge $e_{pc} = \BuildEdge(p_{i}, a_{i})$ \tcp*[l]{add PR-Contributor  edge}
            Edge $e_{pr}= \BuildEdge(p_{i},R_{i})$  \tcp*[l]{add PR-Reviewer  edge}
            $E_{base} \gets \{E_{base}\cup \{e_{pc}\}\cup \{e_{pr}\}\}$\;
        }
        \For{$p_{i} \in \mbox{\it PRS}$}{
            Edge Set $E_i \gets \oslash$\;
            \For{$p_{j} \in \mbox{\it PRS}$} {
                Edge $e_{pp} = \BuildEdge(p_{i}, p_{j})$ \tcp*[l]{add PR-PR edge}
                $E_{i} \gets \{E_{i}\cup \{e_{pp}\}\}$\; 
            }
            $\SortWeightAddFilter(E_{i})$ \tcp*[l]{select high weight edges in set}
            $E_{base} \gets \{E_{base}\cup E_{i}\}$\; 
        }
        $G_{base} \gets \{V_{base},E_{base}\}$\;
        \Return{$G_{base}$} 
    }
    \textbf{EndFunction}
    \BlankLine
    \SetKwFunction{FbuildEdge}{BuildEdge}
    \Fn{\FbuildEdge{$p_{i}, X$}}{
        Edge $e  = \AddVertex(p_{i}, X)$ \tcp*[l]{$X$ can be the contributor($a_{i}$) or Reviewer Set($R_{i}$) or PR($p_{j}$)}
        $w = \CaculateWeight(p_{i}, X)$ \;
        $\SetWeight(e, w)$  \tcp*[l]{add weight to edge}
        \Return{$e$}
    }
    \textbf{EndFunction}
\end{algorithm}

In general, function ~\emph{Construction} describes the main logic for hypergraph construction, which takes review history (contributors, reviewers, \emph{PR}s, etc.) as inputs. Lines 2 initializes the vertex set $V_{base}$ and hyperedge set $E_{base}$. The `for loop' in lines 3-8 updates hypergraph by adding new vertexes of \emph{PR}s, contributors and reviewers as well as hyperedges of \emph{PR-Reviewer} and \emph{PR-Contributor}. The hyperedges representing \emph{PR-PR} relationship should be separately processed (in lines 9-17) since global information is needed to calculate edge weight. Finally, line 19 returns the result hypergraph $G_{base}$. Note that function \emph{Construction} invokes the function~\emph{BuildEdge} to calculate the weights for hyperedges according to different relationships, which is detailed in lines 22-25. Since different types of relationships require different methods to calculate the edge weight, we elaborate them in detail as the following.




\textbf{\emph{PR-Reviewer:}} The relationships between \emph{PR}s and reviewers are necessary for all kinds of graph-based recommenders. In a pull-request development paradigm, one \emph{PR} may experience multiple revisions and re-submissions, which would usually engage multiple reviewers and they may impact each other by publicly posted review comments. Therefore, in addition to the regular relationship, i.e. a pair of one reviewer and one \emph{PR},
reviewers who comment on the same \emph{PR} are connected with a hyperedge in a hypergraph.



In~\emph{HGRec}, the weight of a \emph{PR-Reviewer} edge is set by aggregating all reviewers' contributions, which is formulated in Equation~\ref{EQ:D_PR_REV}. 
\begin{equation}
\label{EQ:D_PR_REV}
    w = \sum_{r_{1_{i}} \in R_{1}} \sum_{j=1}^{o_{1_{i}}} \lambda^{j-1}e^{\frac{t_{i_{j}}-t_{e}}{t_{e}-t_{s}}}
\end{equation}
where reviewers in set $R_{1}$ participated in the \emph{PR} $p_{1}$, reviewer $r_{1_{i}}$ made $o_{1_{i}}$ comments in the \emph{PR} $p_{1}$. The creation time of each comment is $t_{i_{j}}$. The hyperparameter $\lambda$ in Equation~\ref{EQ:D_PR_REV} works for mitigating the influence of comments (cf. subsection \ref{subsec:hyperparameter} for details). Moreover, reviewers' activeness was also considered in~\emph{HGRec}, i.e. the closer reviews are, the greater influence they carried. $t_{s}$ and $t_{e}$ are the start time and the end time of dataset in Equation~\ref{EQ:D_PR_REV}.



\textbf{\emph{PR-Contributor:}} Contributors and reviewers may play different roles in a pull-request development paradigm. Therefore, they are treated differently in~\emph{HGRec} by defining the \textbf{\emph{PR-Contributor}} relationship and the corresponding weight. As Equation~\ref{EQ:D_PR_CO}, the more recent activity is, the higher weight. 
\begin{equation}
\label{EQ:D_PR_CO}
    w = \frac{t_{1}-t_{s}}{t_{e}-t_{s}}
\end{equation}
$t_{1}$ is the creation time of \emph{PR} $p_{1}$. $t_{s}$ and $t_{e}$ take the same meaning as in Equation~\ref{EQ:D_PR_REV}.

\textbf{\emph{PR-PR:}} The profiles (e.g., language, code lines) and content (e.g., the source code) included in \emph{PR}s to a certain degree can reflect the expertise of contributors. Moreover, it is also common that closely located source files share similar functions, and hence can be used for reviewer recommendation~\cite{thongtanunam2015should}.
Therefore, the weight of~\emph{PR-PR} relationship is achieved by considering the distances between~\emph{PR}s in the file path set (as shown in Equation~\ref{EQ:D_PR_PR}).
\begin{equation}
\label{EQ:D_PR_PR}
    w = \sum_{f_{1}\in F_{1},f_{2}\in F_{2}} \frac{\mbox{\it Similarity}(f_{1},f_{2})}{\vert F_{1} \vert \vert F_{2} \vert} e^{-\frac{\vert t_{1}-t_{2}\vert}{t_{e}-t_{s}}}
\end{equation}
where function $\mbox{\it Similarity}$ is calculated as Equation~\ref{EQ:SIMILARITY},

\begin{equation}
\label{EQ:SIMILARITY}
    \mbox{\it Similarity}(f_{1},f_{2}) = \frac{\mbox{\it LCP}(f_{1},f_{2})}{\mbox{\it max}(\mbox{\it len}(f_{1}),\mbox{\it len}(f_{2}))}
\end{equation}
where $F_{1}$, $F_{2}$ are the file path sets contained in two \emph{PR}s, say $p_{1}$ and $p_{2}$.  $f_{1}$ and $f_{2}$ are the specific file paths that belong to the file path sets $F_{1}$ and $F_{2}$.

We also model developers' turnover as an exponential function to smoothen the distance between two \emph{PR}s. In the exponential function of Equation~\ref{EQ:D_PR_PR}, $t_{s}$ and $t_{e}$ are the creation and end time of dataset, $t_{1}$ and $t_{2}$ are the creation time of two \emph{PR}s respectively. Through this way, within a certain time scope, the latest \emph{PR}s are preferentially considered. To reduce calculation cost, we restricted the number of neighbors for a certain \emph{PR} and simplified the hypergraph that only top-$m$~\emph{PR-PR} connections (cf. subsection \ref{subsec:hyperparameter} for details) were included.
Moreover, we employed a MIN-MAX  strategy to normalize the weight for each type of edge.

\subsection{Reviewer Recommendation}
With a constructed hypergraph in hand, we then can perform a recommendation calculation based on the hypergraph. In general, we formulated reviewer recommendation as a ranking task on a hypergraph. Previous studies (e.g.,~\cite{ying2016earec}) used `random walk' strategy to choose neighborhood as the next vertex with a certain probability, which is somehow low-effective. Inspired by~\cite{tan2011using},  we applied an advanced `search and ranking' strategy in \emph{HGRec}, which is elaborated briefly in this subsection.

Given a hypergraph $G_{base}$ and a newly-submitted $p_{t}$, we first develop \emph{PR-PR} relationship by calculating its file path similarities with existing \emph{PR}s in $G_{base}$ and then we connect  $p_{t}$ with the most similar \emph{PR}s. By following a similar strategy (Algorithm \ref{algo_graphConstruction}), we can establish  the \emph{PR-Contributor} relationship. In this way, both new \emph{PR}s and contributors are merged into the original hypergraph $G_{base}$ to form a new  $G_{t}=\{V_{t},E_{t}\}$.

For a hypergraph $G$, the key of this ranking strategy is to find the appropriate ranking vector $f^{*}\in\mathbb{R}^{\vert V\vert}$ which is able to minimize the objective function~$Q(f)$ defined as below:
\begin{equation} \label{EQ:loss}
  \begin{split}
    Q(f)= f^{T}(\bm{I}-\bm{A})f + \mu(f-y)^{T}(f-y)
  \end{split}
\end{equation}
where $y\in\mathbb{R}^{\vert V\vert}$ is a query vector with multiple elements, one for each vertex of the hypergraph $G$ which will be set to 1 for a target \emph{PR} and its contributor, otherwise 0.
$\bm{H}^{G} \in \mathbb{R}^{\vert V\vert \times \vert E \vert} $ is a vertex-hyperedge incidence matrix, $\bm{W^{G}} \in \mathbb{R}^{\vert E\vert \times \vert E \vert}$ is a weight matrix,  $\bm{D}^{G}_{v}$ is a vertex degree matrix and $\bm{D}^{G}_{e}$ is a hyperedge degree matrix,  $\bm{A} = {\bm{D}^G_{v}}^{-1}\bm{H}^{G}\bm{W}^{G}{\bm{D}^{G}_{e}}^{-1}{\bm{H}^{G}}^{T}$, and $\mu$ is the regularization parameter.

Through a series of deductions and transformations, we have the optimal $f^{*}$ as:

\begin{equation} \label{EQ:SOLUTION}
  \begin{split}
    f^{*}&=(\bm{I}-\frac{\bm{A}}{1+\mu})^{-1}y = (I-\alpha \bm{A})^{-1}y
  \end{split}
\end{equation}
where $\alpha = \frac{1}{1 + \mu}$.

Having ranked on the hypergraph, we can recommend the top-$k$ reviewers as the candidates. The whole recommendation process is presented in Algorithm~\ref{algo_graphRecommendation}.

\begin{algorithm}
    \scriptsize
    \caption{Hypergraph-based Recommendation}
    \label{algo_graphRecommendation}
    \SetKwFunction{BuildEdge}{BuildEdge}
    \SetKwFunction{AddVertex}{AddVertex}
    \SetKwFunction{QueryVector}{QueryVector}
    \SetKwFunction{Ranking}{Ranking}
    \SetKwFunction{FilterAndSort}{FilterAndSort}
    \SetKwInOut{Input}{input}\SetKwInOut{Output}{output}
    \Input{Hypergraph $G_{base}$ \\ PR Set $\mbox{\it PRS}$ \\ Target PR $p_{t}$, Contributor $a_{t}$}
    \Output{Recommend List $C_{t}$}
    \BlankLine
    \SetKwFunction{FMain}{Recommendtaion}
    \SetKwProg{Fn}{Function}{}{}
    \Fn{\FMain{$G_{base}, \mbox{\it PRS}, p_{t}, a_{t}$}}{
        $C_{t} \gets \oslash$\;
        $V_{t} \gets \{V_{base}\cup \{p_{t}\}\cup \{a_t\}\}$ \tcp*[l]{add new vertexes to $G_{base}$}
        Edge Set $E_t \gets \oslash$\;
        Edge $e_{pc} = \BuildEdge(p_{t}, a_{t})$ \tcp*[l]{add PR-Contributor edge}
            \For{$p_{i} \in \mbox{\it PRS}$} {
                Edge $e_{pp} = \BuildEdge(p_{t}, p_{i})$ \tcp*[l]{add PR-PR edge}
                $E_{t} \gets \{E_{t}\cup \{e_{pp}\}\}$\; 
            }
        $\SortWeightAddFilter(E_{t})$ \tcp*[l]{select high weight edges in set}
        $E_{t} \gets \{E_{t}\cup \{e_{pc}\}\}$\;
        $G_{t} \gets \{V_{t},E_{t}\}$\;
        $y_{t} \gets \QueryVector(V_{t},p_{t},a_{t})$ \tcp*[l]{use search and ranking strategy}
        $f^{*} \gets \Ranking(G_{t},y_{t})$ \tcp*[l]{get candidates' score}
        $C_{t} \gets \FilterAndSort(f^{*})$ \tcp*[l]{get recommendation list}
        \Return  $C_{t}$
    }
    \textbf{EndFunction}
\end{algorithm}

Algorithm~\ref{algo_graphRecommendation} takes hypergraph $G_{base}$, \emph{PR} set $\mbox{\it PRS}$, and target \emph{PR} as its inputs. Line 2 initializes candidate list $C_{t}$. Line 3 is to add vertexes of three types of entities. Lines 4-12 also invoke function~\textit{BuildEdge} (cf. Algorithm~\ref{algo_graphConstruction}) to build relationships of \emph{PR-Contributor} and \emph{PR-PR}. Line 13 generates the query vector $y_{t}$ by $p_{t}$. Line 14 optimizes objective function $Q(f)$ and get the ranking vector. Lines 15-16 rank and return a recommendation list according to ranking strategy.

\subsection{Hyperparameter Setting}
\label{subsec:hyperparameter}

The hyperparameters, i.e. $\alpha\in[0,1]$, $m\in[5,15]$, $\lambda\in[0,1]$ play critical roles in~\emph{HGRec}. $\alpha$ is the regularization parameter. The smaller $\alpha$, the larger influence of regulation, that is, the weight of vertexes access to query vector $y$. In this case, the nearby reviewers around target \emph{PR}s and contributors have more chances to be recommended. $m$ represents the maximum connections of a \emph{PR}, $\lambda$ represents the influence posed by a reviewer in a history review. Increasing $\alpha$, ~\emph{HGRec} tends to recommend non-core reviewers, which is important to mitigate workload congestion issue discussed in 
Section \ref{Recommendation distribution}. Increasing $m$ or reducing $\lambda$,~\emph{HGRec} tends to recommend core reviewers. Since there are no specific rules to determine $\alpha$, $m$ and $\lambda$, we set them by a `trial-and-error' approach. After many rounds of trial calculations, we found that \emph{HGRec} can produce relatively good results under the following combination of parameters, i.e. $\alpha=0.9$, $m=10$, $\lambda=0.8$.

\section{Evaluation design}
\subsection{Research Questions}
Two research questions (RQs) are proposed for the evaluation, which are
\begin{itemize}
  \item \textbf{RQ1:} To what extent can the proposed~\emph{HGRec} accurately recommend code reviewers?
  \item \textbf{RQ2:} To what extent can the proposed~\emph{HGRec} alleviate workload congestion issue?
\end{itemize}

RQ1 evaluates the performance  of~\emph{HGRec} in terms of accuracy, whereas RQ2 assesses~\emph{HGRec}'s capability of dealing with the other concern -- workload congestion issue.

\subsection{Data Preparation}
GitHub provides multiples APIs to assess various project data. For potential comparison and calibration, the chosen projects are the common ones from previous studies. Since the evaluation involves several time-consuming tasks, as a balance between resources (e.g., time, computing resource,etc.) and the capability to generalize the evaluation results, we selected those projects that appear at least in two of the previous studies containing the baseline recommenders (cf. subsection \ref{comparison recommenders}). As a result, we chose 12 well-known projects in GitHub to evaluate the performance of~\emph{HGRec} as well as other recommenders in order to position our recommender.
The time span of the dataset is from 2017-01-01 to 2020-06-30. Detailed demographics of the dataset are presented in Table~\ref{TAB:project}.

\begin{table}[htp]
\centering
\scriptsize
\caption{Overview of the 12 selected projects}
\label{TAB:project}
    \begin{tabular}{lcccc}
    \toprule
    \textbf{Project} & \textbf{\#PRs} & \textbf{\#Comments} & \textbf{\#Reviewers} & \textbf{\#Contributors} \\
    \midrule
akka                 & 4673           & 45677               & 864                  & 921                     \\
angular           & 12517          & 110178              & 2806                 & 2233                    \\
Baystation12 & 8471           & 41373               & 676                  & 576                     \\
bitcoin           & 7113           & 91092               & 1012                 & 896                     \\
cakephp           & 3319           & 17281               & 860                  & 976                     \\
django             & 6027           & 31607               & 2952                 & 3691                    \\
joomla-cms         & 10327          & 94701               & 2122                 & 1184                    \\
rails               & 7912           & 37720               & 5651                 & 4943                    \\
scala               & 3478           & 24091               & 778                  & 651                     \\
scikit-learn & 6315           & 68903               & 2378                 & 2627                    \\
symfony           & 11283          & 77548               & 3949                 & 3477                    \\
xbmc                 & 5959           & 40451               & 1596                 & 1141                    \\
    \midrule
\textbf{Total}                     & 87394          & 680622              & 25644                & 23316        \\
    \bottomrule
    \end{tabular}
\end{table}

\subsection{Experiment Settings}

\subsubsection{Baselines}
\label{comparison recommenders}
To evaluate ~\emph{HGRec} thoroughly, the following representative traditional recommenders and state-of-the-art recommenders as well are compared as the baselines, which are
\begin{itemize}
    \item \textbf{AC} \cite{jiang2017should} that recommends reviewers based on recent activities of the candidates. Reviewers who leave comments frequently in recent \emph{PR}s are determined to be active and prone to be recommended; otherwise inactive.
    \item \textbf{RevFinder} \cite{thongtanunam2015should} that recommends reviewers by leveraging the file path similarities of \emph{PR}s, i.e. the files located in close files may share similar functionality and therefore should be reviewed by reviewers with similar experience.
    \item \textbf{cHRev} \cite{zanjani2015automatically} that recommends reviewers on the premise that who previously reviewed the code files is tended to be candidate reviewers for a target \emph{PR}.~\emph{cHRev} formulates reviewers' expertise based on ``how many'', ``who performed'', and ``when reviews were performed''.
    \item \textbf{CN} \cite{yu2016reviewer} that recommends reviewers by aggregating developers who share common interests with the contributor of target \emph{PR}.~\emph{CN} mines historical comment traces to construct a comment network to make recommendations.
    \item \textbf{RF} \cite{de2015developers} that recommends reviewers by applying supervised machine learning, i.e. collecting project attributes and \emph{PR}s to construct classifiers and rank candidates.
    \item \textbf{EARec} \cite{ying2016earec} that recommends reviewers by constructing a graph architecture to depict the expertise and authority of developers as well as their interactions. The recommendation is performed using graph searching algorithms.
\end{itemize}

The considerations are three-fold. First, \emph{AC} and \emph{RF} have shown impressively good performance in terms of accuracy in many studies~\cite{de2018automatic, jiang2017should}. Second, \emph{CN} and \emph{EARec} both adopt graph (an ordinary graph) as the underlying model. Last but not least, as two classical recommenders,  \emph{RevFinder} and \emph{cHRev} have been used as the comparison basis frequently in many existing studies.

\subsubsection{Metrics}
To address RQ1, we need to evaluate the performance of \emph{HGRec} in terms of accuracy. We take two common metrics in recommender evaluation studies, i.e. \emph{Accuracy (ACC)} (defined as Equation \ref{EQ:ACC}) and \emph{Mean Reciprocal Rank (MRR)} (defined as Equation \ref{eq:4}).

\begin{itemize}
    \item Accuracy
\end{itemize}

\begin{equation}
\label{EQ:ACC}
    \mbox{\it ACC} = \frac{1}{|\mbox{\it PRS}|} \sum\limits_{p \in \mbox{\it PRS}}\mbox{\it isTrue}(p, k)
\end{equation} 
where, $\mbox{\it PRS}$ is a set of target \emph{PR}s, indicator function~\emph{isTrue(p, k)} returns 1 if the recommended reviewer within top-$k$ candidates finally reviewed the target PR $p$, otherwise returns 0.

\begin{itemize}
    \item Mean Reciprocal Rank
\end{itemize}

\begin{equation}\label{eq:4}
  \mbox{\it MRR} = \frac{1}{|\mbox{\it PRS}|}\sum\limits_{p \in \mbox{\it PRS}}\frac{1}{\mbox{\it rank}(p, k)}
\end{equation}
where function~\emph{rank(p, k)} returns the location where the true reviewer places in the sorted reviewer list. \mbox{\it MRR} rewards score 1  if the first choice was correct and rewards 1/2 if the second choice was correct, and so on. While if the recommended reviewer is not contained in the candidate list, then  \mbox{\it MRR} rewards 0. The final \emph{MRR} is calculated as the average value of all the scores.

To answer RQ2, we defined  \emph{Recommendation Distribution (RD)} as Equation \ref{EQ:Entropy} to measure the extent that diverse reviewers can be recommended.

\begin{itemize}
    \item Recommendation Distribution (RD)
\end{itemize}

\begin{equation}
\label{EQ:Entropy}
  {\mbox{\it RD} = -\frac{1}{log_2 n} \sum_{i=1}^n P(i) log_2 P(i)}
\end{equation}
where,~\emph{n} is the total number of reviewers, $P(i)$ is a percentage that indicates the workload of the $i_{th}$ reviewer. The larger \emph{RD}, the more diverse that a recommender recommends reviewers.

As a popular standard in the related studies, we evaluated the top-$k$ ($k$=1, 3, 5) performances of the recommenders. To further test the difference, we established hypotheses and applied the Wilcoxon Signed Rank Test on \emph{ACC}, \emph{MRR} and \emph{RD}. The null and alternative hypotheses can be stated as follows,

  \begin{description}
    \item[$H_{0,M}$]: \emph{There is no significant difference on the metrics M between HGRec and  R}.
    \item[$H_{1a,M}$]: \emph{HGRec is significantly better than R on metrics M}.
    \item[$H_{1b,M}$]: \emph{HGRec is significantly worse than R on metrics M}.
  \end{description}
  where  \emph{M}  can be \emph{ACC}, \emph{MRR} and \emph{RD} and correspondingly,  \emph{R} represents one recommender introduced in Section~\ref{comparison recommenders}.

\subsubsection{Data pre-processing}
\label{subsec:data_preocessing}
\begin{figure}[!htbp]
  \centering
  \includegraphics[width=1.0\linewidth]{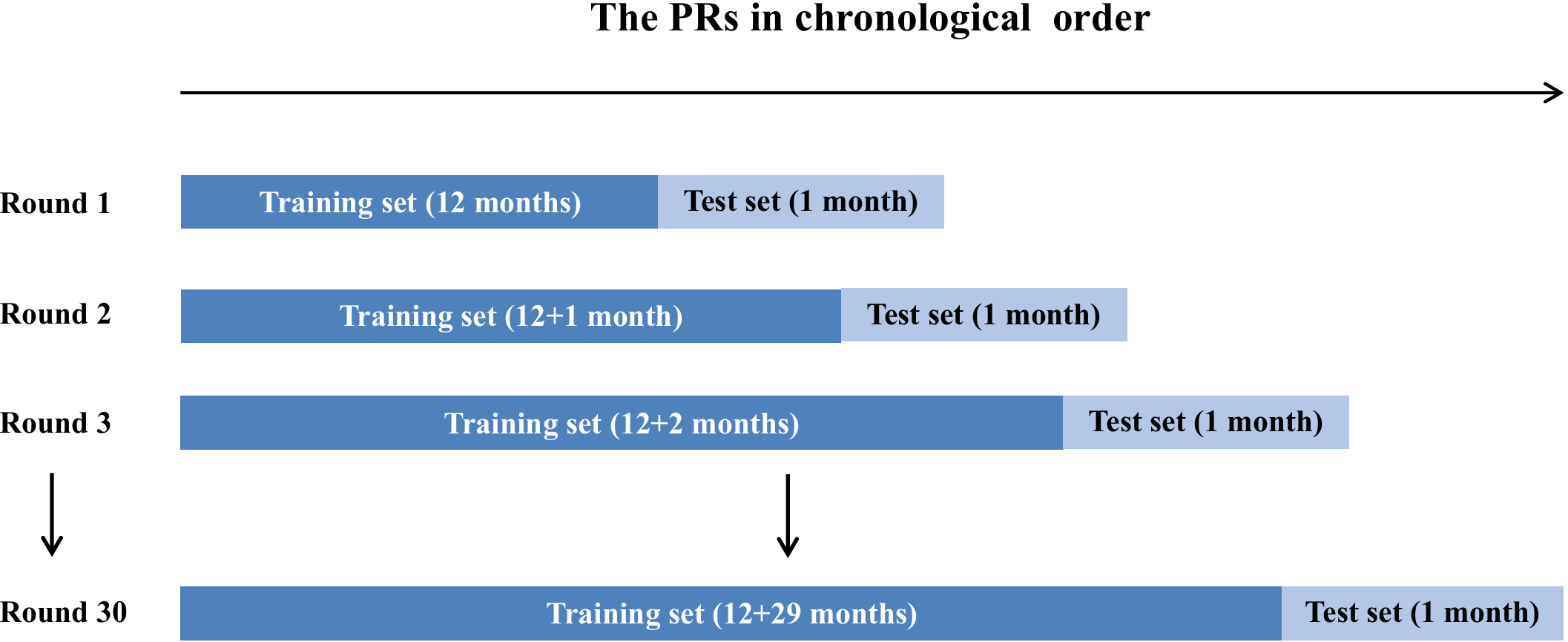}
  \caption{Dataset setting for evaluation}
  \label{FIG:stepOfdataset}
\end{figure}

Following the similar method in ~\cite{lipcak2018large, de2018automatic}, we applied a time series strategy to evaluate the recommenders' performance in terms of \emph{ACC}, \emph{MRR} and \emph{RD}. To be specific, all the reviews in 2017 were initiated as the original training set and hereafter each monthly review until Jun, 2020 played the role of the test set. Therefore, we eventually performed  30 rounds of evaluation in total, as shown in  Figure~\ref{FIG:stepOfdataset}. Take the first round for example, the first 12-month data is fed into all the recommenders and then the data of the 13th month is used to calculate \emph{ACC}, \emph{MRR} and \emph{RD} using Equation \ref{EQ:ACC}, \ref{eq:4} and \ref{EQ:Entropy}, respectively.

\begin{table*}[t!hp]
\caption{\emph{ACC} of recommenders}
    \label{tab:top-K_results}
\scriptsize
\begin{threeparttable}
    \begin{tabular}{p{0.3cm}p{0.3cm}p{0.3cm}p{0.4cm}p{0.3cm}p{0.3cm}p{0.4cm}p{0.3cm}p{0.3cm}p{0.4cm}p{0.3cm}p{0.3cm}p{0.4cm}p{0.3cm}p{0.3cm}p{0.4cm}p{0.3cm}p{0.3cm}p{0.4cm}p{0.3cm}p{0.3cm}p{0.4cm}p{0.3cm}p{0.3cm}p{0.4cm}}
\toprule
                                  \multicolumn{1}{l}{}  & \multicolumn{3}{c}{RevFinder} & \multicolumn{3}{c}{CN} & \multicolumn{3}{c}{AC}                          & \multicolumn{3}{c}{cHRev} & \multicolumn{3}{c}{RF}                 & \multicolumn{3}{c}{EARec} & \multicolumn{3}{c}{HGRec}                       \\ \hline
\multicolumn{1}{l}{}             & 1        & 3        & 5        & 1      & 3      & 5     & 1              & 3              & 5              & 1       & 3       & 5      & 1              & 3              & 5     & 1       & 3       & 5      & 1              & 3              & 5              \\ \hline
\multicolumn{1}{l}{akka}             & 0.515    & 0.893    & 0.976   & 0.614  & 0.935 & 0.980 & 0.531          & 0.922          & 0.986          & 0.480   & 0.880  & 0.967  & 0.612          & 0.948          & 0.984 & 0.486   & 0.876  & 0.968  & \textbf{0.638} & \textbf{0.950} & \textbf{0.987} \\
\multicolumn{1}{l}{angular}           & 0.298    & 0.567    & 0.719   & 0.401  & 0.655 & 0.762 & 0.215          & 0.514          & 0.666          & 0.385   & 0.674  & 0.784  & 0.292          & 0.569          & 0.688 & 0.176   & 0.456  & 0.628  & \textbf{0.483} & \textbf{0.744} & \textbf{0.834} \\
\multicolumn{1}{l}{Baystation12}      & 0.334    & 0.639    & 0.765   & 0.295  & 0.707 & 0.820 & 0.393          & \textbf{0.811} & \textbf{0.926} & 0.366   & 0.704  & 0.824  & \textbf{0.399} & 0.684          & 0.792 & 0.326   & 0.645  & 0.731  & 0.373          & 0.756          & 0.873          \\
\multicolumn{1}{l}{bitcoin}            & 0.490    & 0.819    & 0.907   & 0.536  & 0.813 & 0.904 & 0.493          & 0.842          & 0.921          & 0.409   & 0.751  & 0.872  & 0.516          & 0.852          & 0.924 & 0.472   & 0.815  & 0.887  & \textbf{0.574} & \textbf{0.859} & \textbf{0.930} \\
\multicolumn{1}{l}{cakephp}         & 0.529    & 0.860    & 0.920   & 0.576  & 0.881 & 0.948 & 0.527          & 0.870          & 0.957          & 0.492   & 0.826  & 0.919  & 0.585          & \textbf{0.910} & 0.950 & 0.526   & 0.870  & 0.918  & \textbf{0.587} & 0.903          & \textbf{0.961} \\
\multicolumn{1}{l}{django}       & 0.289    & 0.781    & 0.898   & 0.429  & 0.806 & 0.855 & \textbf{0.595} & \textbf{0.834} & \textbf{0.917} & 0.483   & 0.741  & 0.834  & 0.376          & 0.792          & 0.910 & 0.285   & 0.744  & 0.898  & 0.564          & 0.808          & 0.881          \\
\multicolumn{1}{l}{joomla-cms}    & 0.475    & 0.727    & 0.825   & 0.503  & 0.752 & 0.844 & 0.500          & 0.743          & 0.853          & 0.313   & 0.616  & 0.760  & 0.492          & 0.735          & 0.833 & 0.407   & 0.700  & 0.804  & \textbf{0.532} & \textbf{0.772} & \textbf{0.863} \\
\multicolumn{1}{l}{rails}            & 0.294    & 0.526    & 0.656   & 0.266  & 0.481 & 0.594 & 0.294          & \textbf{0.540} & \textbf{0.660} & 0.251   & 0.460  & 0.576  & \textbf{0.297} & 0.500          & 0.616 & 0.294   & 0.450  & 0.622  & 0.296          & 0.532          & 0.651          \\
\multicolumn{1}{l}{scala}            & 0.358    & 0.625    & 0.785   & 0.371  & 0.692 & 0.810 & 0.368          & 0.645          & 0.790          & 0.307   & 0.643  & 0.781  & 0.382          & 0.658          & 0.784 & 0.261   & 0.597  & 0.758  & \textbf{0.413} & \textbf{0.714} & \textbf{0.831} \\
\multicolumn{1}{l}{scikit-learn}    & 0.484    & 0.666    & 0.798   & 0.484  & 0.707 & 0.851 & \textbf{0.510} & \textbf{0.796} & \textbf{0.913} & 0.408   & 0.682  & 0.796  & 0.499          & 0.725          & 0.856 & 0.483   & 0.616  & 0.754  & 0.505          & 0.758          & 0.877          \\
\multicolumn{1}{l}{symfony}   & 0.529    & 0.898    & 0.940   & 0.607  & 0.888 & 0.932 & 0.560          & 0.900          & 0.938          & 0.514   & 0.829  & 0.912  & 0.619          & 0.908          & 0.939 & 0.524   & 0.898  & 0.926  & \textbf{0.626} & \textbf{0.916} & \textbf{0.947} \\
\multicolumn{1}{l}{xbmc}  & 0.229    & 0.460    & 0.604   & 0.289  & 0.538 & 0.669 & 0.268          & 0.477          & 0.573          & 0.290   & 0.561  & 0.680  & 0.263          & 0.473          & 0.582 & 0.215   & 0.393  & 0.489  & \textbf{0.367} & \textbf{0.632} & \textbf{0.740} \\
\hline
\multicolumn{1}{l}{AVG}  & 0.402    & 0.705    & 0.816   & 0.447  & 0.738 & 0.831 & 0.438          & 0.741          & 0.842          & 0.391   & 0.697  & 0.809  & 0.444          & 0.729          & 0.821 & 0.371   & 0.672  & 0.782  & \textbf{0.496} & \textbf{0.779} & \textbf{0.865}     \\ 
\bottomrule
\end{tabular}

\end{threeparttable}
\end{table*}

\begin{table*}[t!hp]
\caption{\emph{MRR} of recommenders}
    \label{tab:mrr_results}
\scriptsize
\begin{threeparttable}
    \begin{tabular}{p{0.3cm}p{0.3cm}p{0.3cm}p{0.4cm}p{0.3cm}p{0.3cm}p{0.4cm}p{0.3cm}p{0.3cm}p{0.4cm}p{0.3cm}p{0.3cm}p{0.4cm}p{0.3cm}p{0.3cm}p{0.4cm}p{0.3cm}p{0.3cm}p{0.4cm}p{0.3cm}p{0.3cm}p{0.4cm}p{0.3cm}p{0.3cm}p{0.4cm}}
\toprule
                                  \multicolumn{1}{l}{}  & \multicolumn{3}{c}{RevFinder} & \multicolumn{3}{c}{CN} & \multicolumn{3}{c}{AC}                          & \multicolumn{3}{c}{cHREV} & \multicolumn{3}{c}{RF}                 & \multicolumn{3}{c}{EARec} & \multicolumn{3}{c}{HGRec}                       \\ \hline
           \multicolumn{1}{l}{}             & 1        & 3        & 5        & 1      & 3      & 5     & 1              & 3              & 5              & 1       & 3       & 5      & 1              & 3              & 5     & 1       & 3       & 5      & 1              & 3              & 5              \\ \hline
\multicolumn{1}{l}{akka}   & 0.515    & 0.689    & 0.708   & 0.614  & 0.760 & 0.771 & 0.531          & 0.704          & 0.719          & 0.480   & 0.657  & 0.678  & 0.612          & 0.764          & 0.772 & 0.486   & 0.666  & 0.687  & \textbf{0.638} & \textbf{0.779} & \textbf{0.787} \\
\multicolumn{1}{l}{angular}      & 0.298    & 0.414    & 0.449   & 0.401  & 0.512 & 0.536 & 0.215          & 0.344          & 0.378          & 0.385   & 0.512  & 0.538  & 0.292          & 0.412          & 0.440 & 0.176   & 0.296  & 0.336  & \textbf{0.483} & \textbf{0.598} & \textbf{0.619} \\
\multicolumn{1}{l}{Baystation12}   & 0.334    & 0.473    & 0.502   & 0.295  & 0.475 & 0.501 & 0.393          & \textbf{0.577} & \textbf{0.604} & 0.366   & 0.515  & 0.542  & \textbf{0.399} & 0.531          & 0.555 & 0.326   & 0.476  & 0.495  & 0.373          & 0.542          & 0.569          \\
\multicolumn{1}{l}{bitcoin}       & 0.490    & 0.637    & 0.658   & 0.536  & 0.658 & 0.679 & 0.493          & 0.650          & 0.668          & 0.409   & 0.559  & 0.586  & 0.516          & 0.668          & 0.684 & 0.472   & 0.625  & 0.642  & \textbf{0.574} & \textbf{0.703} & \textbf{0.719} \\
\multicolumn{1}{l}{cakephp}    & 0.529    & 0.672    & 0.686   & 0.576  & 0.716 & 0.731 & 0.527          & 0.674          & 0.695          & 0.492   & 0.638  & 0.660  & 0.585          & \textbf{0.732} & 0.741 & 0.526   & 0.672  & 0.684  & \textbf{0.587} & 0.729          & \textbf{0.743} \\
\multicolumn{1}{l}{django}    & 0.289    & 0.508    & 0.535   & 0.429  & 0.596 & 0.607 & \textbf{0.595} & \textbf{0.699} & \textbf{0.718} & 0.483   & 0.597  & 0.619  & 0.376          & 0.569          & 0.597 & 0.285   & 0.503  & 0.540  & 0.564          & 0.671          & 0.688          \\
\multicolumn{1}{l}{joomla-cms}   & 0.475    & 0.589    & 0.611   & 0.503  & 0.613 & 0.634 & 0.500          & 0.606          & 0.632          & 0.313   & 0.444  & 0.477  & 0.492          & 0.603          & 0.625 & 0.407   & 0.543  & 0.566  & \textbf{0.532} & \textbf{0.638} & \textbf{0.659} \\
\multicolumn{1}{l}{rails}   & 0.294    & 0.391    & 0.420   & 0.266  & 0.359 & 0.385 & 0.294          & 0.397          & 0.424          & 0.251   & 0.340  & 0.367  & \textbf{0.297} & 0.383          & 0.409 & 0.294   & 0.355  & 0.396  & 0.296          & \textbf{0.399} & \textbf{0.426} \\
\multicolumn{1}{l}{scala}     & 0.358    & 0.476    & 0.513   & 0.371  & 0.511 & 0.538 & 0.368          & 0.488          & 0.521          & 0.307   & 0.453  & 0.485  & 0.382          & 0.504          & 0.532 & 0.261   & 0.417  & 0.455  & \textbf{0.413} & \textbf{0.544} & \textbf{0.571} \\
\multicolumn{1}{l}{scikit-learn}   & 0.484    & 0.560    & 0.590   & 0.484  & 0.583 & 0.616 & \textbf{0.510} & \textbf{0.631} & \textbf{0.659} & 0.408   & 0.527  & 0.553  & 0.499          & 0.596          & 0.626 & 0.483   & 0.538  & 0.570  & 0.505          & 0.615          & 0.642          \\
\multicolumn{1}{l}{symfony}   & 0.529    & 0.708    & 0.718   & 0.607  & 0.741 & 0.751 & 0.560          & 0.724          & 0.733          & 0.514   & 0.655  & 0.675  & 0.619          & 0.759          & 0.766 & 0.524   & 0.706  & 0.712  & \textbf{0.626} & \textbf{0.763} & \textbf{0.770} \\
\multicolumn{1}{l}{xbmc}   & 0.229    & 0.331    & 0.363   & 0.289  & 0.395 & 0.425 & 0.268          & 0.358          & 0.379          & 0.290   & 0.408  & 0.435  & 0.263          & 0.353          & 0.378 & 0.215   & 0.288  & 0.310  & \textbf{0.367} & \textbf{0.483} & \textbf{0.507} \\
\hline
\multicolumn{1}{l}{AVG}  & 0.402    & 0.537    & 0.563   & 0.447  & 0.577 & 0.598 & 0.438          & 0.571          & 0.594          & 0.391   & 0.526  & 0.551  & 0.444          & 0.573          & 0.594 & 0.371   & 0.507  & 0.533  & \textbf{0.496} & \textbf{0.622} & \textbf{0.642}                \\ 
\bottomrule
\end{tabular}

\end{threeparttable}
\end{table*}

\begin{table*}[!thp]
\caption{\emph{RD} of recommenders}
    \label{tab:entropy_results}
\scriptsize
\begin{threeparttable}
    \begin{tabular}{p{0.3cm}p{0.3cm}p{0.3cm}p{0.4cm}p{0.3cm}p{0.3cm}p{0.4cm}p{0.3cm}p{0.3cm}p{0.4cm}p{0.3cm}p{0.3cm}p{0.4cm}p{0.3cm}p{0.3cm}p{0.4cm}p{0.3cm}p{0.3cm}p{0.4cm}p{0.3cm}p{0.3cm}p{0.4cm}p{0.3cm}p{0.3cm}p{0.4cm}}
\toprule
                                  \multicolumn{1}{l}{}  & \multicolumn{3}{c}{RevFinder} & \multicolumn{3}{c}{CN} & \multicolumn{3}{c}{AC}                          & \multicolumn{3}{c}{cHRev} & \multicolumn{3}{c}{RF}                 & \multicolumn{3}{c}{EARec} & \multicolumn{3}{c}{HGRec}                       \\ \hline
           \multicolumn{1}{l}{}             & 1        & 3        & 5        & 1      & 3      & 5     & 1              & 3              & 5              & 1       & 3       & 5      & 1              & 3              & 5     & 1       & 3       & 5      & 1              & 3              & 5              \\ \hline
\multicolumn{1}{l}{akka}     & 0.090    & 0.200    & 0.286   & 0.153  & 0.286 & 0.352 & 0.030  & 0.195 & 0.284 & \textbf{0.232} & \textbf{0.307} & \textbf{0.366} & 0.122  & 0.243 & 0.309 & 0.003   & 0.197  & 0.286  & 0.182   & 0.278  & 0.356  \\
\multicolumn{1}{l}{angular}     & 0.150    & 0.241    & 0.318   & 0.282  & 0.355 & 0.403 & 0.024  & 0.170 & 0.236 & \textbf{0.345} & \textbf{0.399} & \textbf{0.438} & 0.170  & 0.214 & 0.275 & 0.002   & 0.159  & 0.229  & 0.305   & 0.372  & 0.413  \\
\multicolumn{1}{l}{Baystation12}    & 0.034    & 0.211    & 0.297   & 0.243  & 0.337 & 0.411 & 0.027  & 0.205 & 0.287 & \textbf{0.265} & \textbf{0.363} & \textbf{0.449} & 0.088  & 0.237 & 0.316 & 0.016   & 0.195  & 0.282  & 0.215   & 0.327  & 0.415  \\
\multicolumn{1}{l}{bitcoin}   & 0.076    & 0.224    & 0.301   & 0.204  & 0.306 & 0.371 & 0.012  & 0.171 & 0.256 & \textbf{0.315} & \textbf{0.386} & \textbf{0.416} & 0.061  & 0.225 & 0.289 & 0.000   & 0.171  & 0.250  & 0.198   & 0.283  & 0.351  \\
\multicolumn{1}{l}{cakephp}     & 0.022    & 0.226    & 0.321   & 0.134  & 0.296 & 0.377 & 0.005  & 0.207 & 0.304 & \textbf{0.182} & \textbf{0.308} & \textbf{0.394} & 0.079  & 0.263 & 0.325 & 0.000   & 0.206  & 0.307  & 0.138   & 0.295  & 0.388  \\
\multicolumn{1}{l}{django}          & 0.005    & 0.201    & 0.254   & 0.108  & 0.263 & 0.332 & 0.005  & 0.178 & 0.245 & \textbf{0.153} & \textbf{0.313} & \textbf{0.391} & 0.061  & 0.212 & 0.262 & 0.000   & 0.164  & 0.243  & 0.099   & 0.299  & 0.370  \\
\multicolumn{1}{l}{joomla-cms}   & 0.119    & 0.240    & 0.294   & 0.172  & 0.310 & 0.367 & 0.004  & 0.176 & 0.253 & \textbf{0.354} & \textbf{0.429} & \textbf{0.468} & 0.088  & 0.219 & 0.287 & 0.001   & 0.166  & 0.244  & 0.146   & 0.309  & 0.373  \\
\multicolumn{1}{l}{rails}      & 0.067    & 0.201    & 0.260   & 0.204  & 0.292 & 0.350 & 0.004  & 0.158 & 0.228 & \textbf{0.317} & \textbf{0.391} & \textbf{0.436} & 0.010  & 0.182 & 0.238 & 0.000   & 0.150  & 0.223  & 0.243   & 0.333  & 0.392  \\
\multicolumn{1}{l}{scala}  & 0.032    & 0.240    & 0.329   & 0.271  & 0.350 & 0.417 & 0.015  & 0.212 & 0.295 & \textbf{0.306} & \textbf{0.395} & \textbf{0.445} & 0.067  & 0.242 & 0.334 & 0.029   & 0.197  & 0.288  & 0.258   & 0.355  & 0.423  \\
\multicolumn{1}{l}{scikit-learn}      & 0.011    & 0.191    & 0.272   & 0.113  & 0.276 & 0.333 & 0.008  & 0.179 & 0.255 & \textbf{0.238} & \textbf{0.349} & \textbf{0.393} & 0.042  & 0.223 & 0.297 & 0.000   & 0.169  & 0.251  & 0.150   & 0.308  & 0.361  \\
\multicolumn{1}{l}{symfony}     & 0.021    & 0.159    & 0.230   & 0.121  & 0.266 & 0.338 & 0.017  & 0.165 & 0.234 & \textbf{0.196} & \textbf{0.322} & \textbf{0.395} & 0.087  & 0.209 & 0.249 & 0.000   & 0.154  & 0.226  & 0.114   & 0.273  & 0.365  \\
\multicolumn{1}{l}{xbmc}       & 0.145    & 0.240    & 0.326   & 0.296  & 0.387 & 0.466 & 0.017  & 0.200 & 0.284 & \textbf{0.391} & \textbf{0.470} & \textbf{0.513} & 0.156  & 0.245 & 0.315 & 0.000   & 0.181  & 0.270  & 0.281   & 0.413  & 0.483  \\
\hline
\multicolumn{1}{l}{AVG}      & 0.064    & 0.214    & 0.291   & 0.192  & 0.310 & 0.376 & 0.014  & 0.185 & 0.263 & \textbf{0.274} & \textbf{0.369} & \textbf{0.425} & 0.086  & 0.226 & 0.291 & 0.004   & 0.176  & 0.258  & 0.194   & 0.321  & 0.391    \\ 
\bottomrule
\end{tabular}

\end{threeparttable}
\end{table*}

\begin{table}[!htp]
\caption{Number of projects by Wilcoxon Signed Rank Test on \emph{ACC} \& \emph{MRR} \& \emph{RD} of recommenders}
    \label{tab:top-K_results_t_test_all}
\scriptsize
\begin{threeparttable}
    \begin{tabular}{p{0.8cm}p{0.25cm}p{0.4cm}p{0.25cm}p{0.45cm}p{0.4cm}p{0.25cm}p{0.45cm}p{0.4cm}p{0.25cm}p{0.45cm}}
\toprule
                                  \multicolumn{2}{c}{}  & \multicolumn{3}{c}{ACC} & \multicolumn{3}{c}{MRR} & \multicolumn{3}{c}{RD}                              \\ \hline

\multicolumn{1}{l}{$M$}      &\multicolumn{1}{l}{top-k}   &  $H_{1a,M}$     & $H_{0,M}$& $H_{1b,M}$    &  $H_{1a,M}$     & $H_{0,M}$& $H_{1b,M}$     &  $H_{1a,M}$     & $H_{0,M}$& $H_{1b,M}$  \\
\midrule
\multirow{3}{*}{RevFinder} & 1 & 9       & 3     & 0     & 9       & 3     & 0     & 12     & 0     & 0     \\
                           & 3 & 10      & 2     & 0     & 11      & 1     & 0     & 12     & 0     & 0     \\
                           & 5 & 10      & 1     & 1     & 11      & 1     & 0     & 12     & 0     & 0     \\
\multirow{3}{*}{CN}        & 1 & 9       & 3     & 0     & 9       & 3     & 0     & 4      & 5     & 3     \\
                           & 3 & 10      & 2     & 0     & 11      & 1     & 0     & 6      & 3     & 3     \\
                           & 5 & 11      & 1     & 0     & 11      & 1     & 0     & 9      & 2     & 1     \\
\multirow{3}{*}{AC}        & 1 & 7       & 5     & 0     & 7       & 5     & 0     & 12     & 0     & 0     \\
                           & 3 & 8       & 1     & 3     & 8       & 2     & 2     & 12     & 0     & 0     \\
                           & 5 & 5       & 4     & 3     & 8       & 1     & 3     & 12     & 0     & 0     \\
\multirow{3}{*}{cHRev}     & 1 & 11      & 1     & 0     & 11      & 1     & 0     & 0      & 0     & 12    \\
                           & 3 & 12      & 0     & 0     & 11      & 1     & 0     & 0      & 0     & 12    \\
                           & 5 & 12      & 0     & 0     & 11      & 1     & 0     & 0      & 1     & 11    \\
\multirow{3}{*}{RF}        & 1 & 5       & 7     & 0     & 5       & 7     & 0     & 12     & 0     & 0     \\
                           & 3 & 6       & 6     & 0     & 6       & 6     & 0     & 12     & 0     & 0     \\
                           & 5 & 8       & 3     & 1     & 6       & 6     & 0     & 12     & 0     & 0     \\
\multirow{3}{*}{EARec}     & 1 & 9       & 3     & 0     & 9       & 3     & 0     & 12     & 0     & 0     \\
                           & 3 & 12      & 0     & 0     & 12      & 0     & 0     & 12     & 0     & 0     \\
                           & 5 & 10      & 1     & 1     & 12      & 0     & 0     & 12     & 0     & 0      \\
\bottomrule
\end{tabular}
\end{threeparttable}
\vspace{-0.6cm}
\end{table}

\section{Results and analysis}

Following a common strategy~\cite{thongtanunam2015should, yu2016reviewer, zanjani2015automatically}, we compare \emph{HGRec} with other recommenders in terms of \emph{ACC}, \emph{MRR} and \emph{RD} using top-1, top-3 and top-5 criteria, respectively. This section presents the results and the corresponding analysis.
\subsection{Accuracy (RQ1)}
\label{sub:accuracy}

Table~\ref{tab:top-K_results} shows the performance of each recommender in terms of \emph{ACC}. The results in bold mean the best recommender regarding \emph{ACC} for a certain project. For example, ~\emph{HGRec} performed the best in project `akka' with all the top-1, top-3 and top-5 criteria. In general,~\emph{HGRec},~\emph{AC} and~\emph{RF} performed relatively better than other recommenders in most cases. To be specific,~\emph{HGRec} takes the lead on 8 (7, 8) projects in terms of top-1 (top-3, top-5) \emph{ACC}. As the close competitors, \emph{AC} wins on 2 (4,4) projects, ~\emph{RF} leads on 2 (1,0) projects using the same top-1, (top-3, top-5) criteria. The last row in Table ~\ref{tab:top-K_results} lists the average \emph{ACC} for all the recommenders, which further indicates \emph{HGRec}'s superiority. With all the top-1, (top-3, top-5) criteria, \emph{HGRec} produces the best average \emph{ACC}. Besides, compared with other two recommenders (i.e. \emph{CN} and \emph{RARec}) using  graph techniques, \emph{HGRec} outperforms the others regarding \emph{ACC} for both solo project and the overall average, indicating the advantage of hypergraph technique to model the review history. Moreover, as the comparison basis, the \emph{ACC} given by recommender \emph{RevFinder} and \emph{cHRev} is not ideal, which to a fair degree is in line with other studies~\cite{yu2016reviewer, jiang2017should, mirsaeedi2020mitigating}. To further test the difference,  a Wilcoxon Signed Rank Test has been conducted on \emph{ACC} using the data from all the 12 projects. Note that there are 30 data points in each project according to the experimental setting elaborated in subsection \ref{subsec:data_preocessing}. Due to page limits \footnote{The dataset, source code and complete results are now public online through https://doi.org/10.6084/m9.figshare.19199981.v1}, we present the number of projects in which we are not able to reject a certain hypothesis (i.e.,$H_{0,M}$, $H_{1a,M}$ and $H_{1a,M}$, where \emph{M} denotes \emph{ACC}) with \emph{p-value} 0.05. The results are listed in Table \ref{tab:top-K_results_t_test_all} (the 3 columns under \emph{ACC}). Take the first row as an example, in 9 out of 12 projects, \emph{HGRec} produces a significantly better \emph{ACC} than recommender \emph{RevFinder} using the top-1 criteria, meanwhile, there are 3 projects in which no significant difference on \emph{ACC} between \emph{RevFinder} and \emph{HGRec} has been observed. The rest is similar, which also confirms our intuitive observation derived from Table \ref{tab:top-K_results}, i.e. \emph{HGRec} performed the best on \emph{ACC} among the recommenders involved in this study.

\emph{MRR} also measures the performance regarding recommendation accuracy. Based on Table \ref{tab:mrr_results} and Table \ref{tab:top-K_results_t_test_all}, we are able to observe similar results, i.e. \emph{HGRec} leads the performance on recommendation accuracy among all the recommenders mentioned in this study.

\begin{figure*}[htp]
\centering
\includegraphics[width=0.95\linewidth]{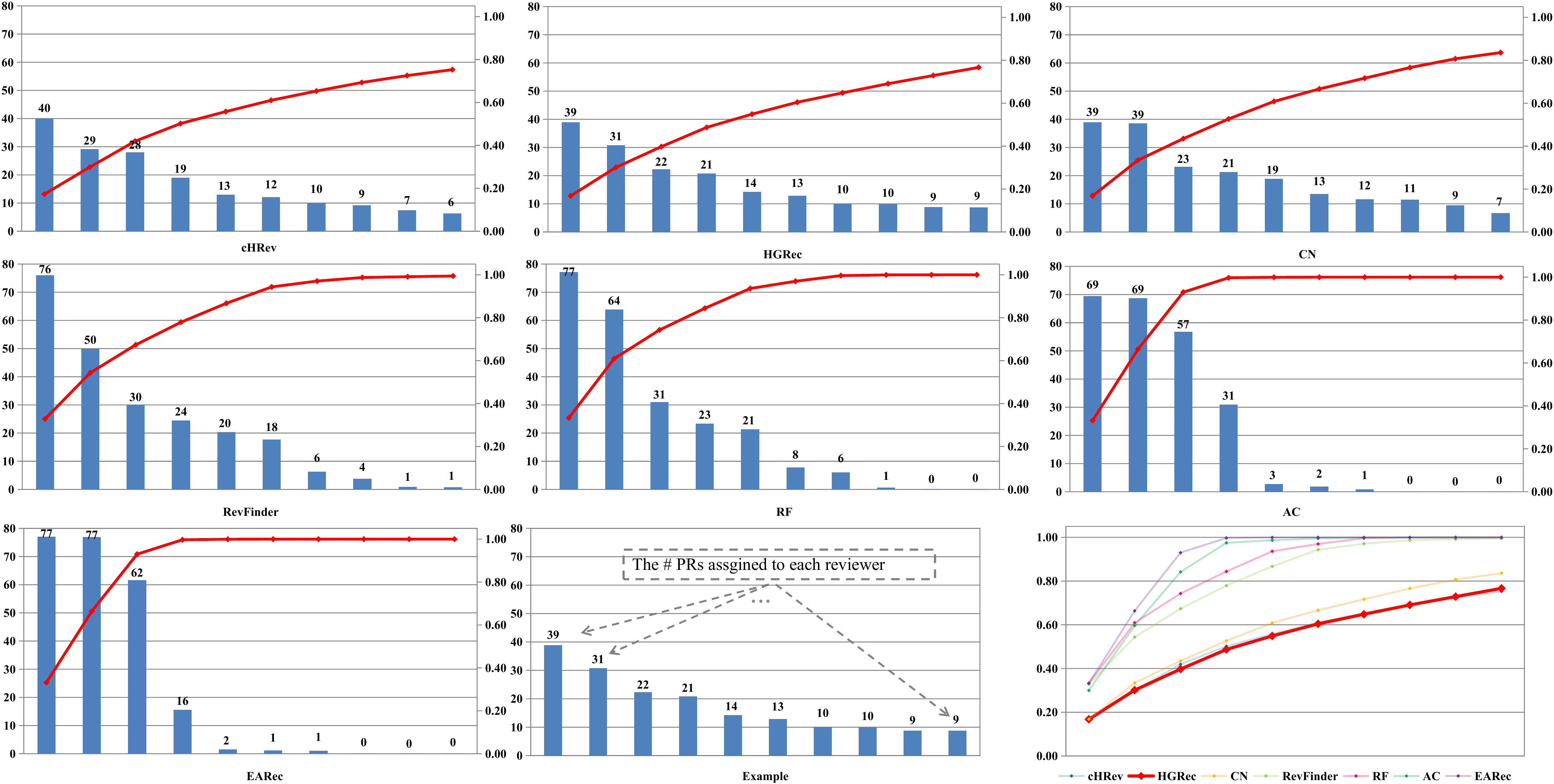}
\caption{Top-3 workload distributions of recommenders on project `angular' (recommended reviewers in $X$-axis, \# \emph{PR}s assigned to each reviewer in $Y$-axis)}
\label{fig:recommendation_distribution}
\end{figure*}

\subsection{Workload (RQ2)}

Recently, researchers raise a new concern other than the accuracy of recommenders. That is, most recommenders tend to suggest a small group of core reviewers (i.e. the reviewers who reviewed the most \emph{PR}s in a certain project have more chance to be recommended). This phenomenon may cause severe problems in projects with massive \emph{PR}s within a relatively short period. \emph{RD} is quantified to show the status of workload congestion.

Similarly, we also evaluated recommenders' performance regarding \emph{RD} and further tested the difference using Wilcoxon Signed Rank Test. Table \ref{tab:entropy_results} lists the results of \emph{RD} for all the recommenders on all the 12 projects. In general, \emph{cHRev} performs the best in all the recommenders. Meanwhile, next to \emph{cHRev}, \emph{HGRec} and \emph{CN} present comparable performance regarding \emph{RD}. Nevertheless, \emph{cHRev} and \emph{CN} showed relatively poor performance regarding \emph{ACC}.
A noteworthy point is that among the best 3 recommenders regarding \emph{ACC}, \emph{HGRec} outperforms the others, i.e. \emph{RF} and \emph{AC} by a discernible margin. The  Wilcoxon Signed Rank Test results listed in Table~\ref{tab:top-K_results_t_test_all} also confirm these observations.



To present an intuitive concept, take project `angular' as an example, Figure~\ref{fig:recommendation_distribution} shows the workload distribution resulting from diverse recommenders based on the number of \emph{PR}s in one month for project `angular'.
Each column represents one reviewer's workload (i.e. \# \emph{PR}s assigned), the broken line represents accumulated workload on diverse reviewers. For top-3 accurate recommenders, ~\emph{HGRec} tends to create a relatively balanced workload for top-10 core reviewers. Take ~\emph{RF} for example, the top-2 core reviewers are recommended for reviewing 77 and 64 \emph{PR}s in just one month, which might be huge burdens for them. As a comparison, the  top-2 core reviewers are recommended by~\emph{HGRec} for merely 39 and 31 \emph{PR}s. 

The reason behind this phenomenon is that by properly tuning $\alpha$ and $\lambda$ in \emph{HGRec}, the importance (score) for some reviewers sharing the review experience on the same \emph{PR}s has been increased, which increases their chances to be recommended, even they may not be active reviewers in the past.

\section{Discussion}
\subsection{Graph Technology for Recommenders}
To represent relationships in a review recommendation paradigm is a basis of a recommender. Although some graph technologies have been adopted to model the relationships when developing recommenders, the primary innovation in ~\emph{HGRec} lies in the introduction of hypergraph to model multiple participants involved in one \emph{PR}, which is very common in OSS projects and easy to understand. Compared with the traditional recommenders that are based on the ordinary graphs,~\emph{HGRec} consists of multiple vertexes and hyperedges that can naturally model the complex high-order relationships among \emph{PR}s, contributors and reviewers.
 Besides,~\emph{HGRec} supports a flexible recommendation architecture, that is more entities and relationships, e.g., organization, comments can be involved in~\emph{HGRec} if necessary.

When it comes to recommenders with similar technologies,~\emph{CN} only considers simple relationships such as developer vertexes, their interactive relationships (e.g., review activities) are formulated by directed edges.~\emph{CN} suggests candidate reviewers who share common interest with contributors but neglects the \emph{PR} information itself.
On the contrary,~\emph{EARec} includes both \emph{PR} and reviewer vertexes, and recommends candidates by matching the characteristics of target \emph{PR} and expertise of candidates. However,~\emph{EARec} does not consider the information of contributors' internal relationships, e.g., the social relationships, which may not be able to be directly obtained from reviewer's profile.~\emph{HGRec} systematically combines multiple roles, including \emph{PR}s' content and interactive relationships among the three entities (i.e. \emph{PR}s, contributors and reviewers). More importantly, ~\emph{HGRec} is able to model the complex in an intuitive manner close to the reality.

\vspace{-0.2cm}
\subsection{Model Interpretability}
In recent years, AI (Artificial Intelligence )/ML (Machine Learning)  techniques are widely used in software engineering, such as software defect prediction~\cite{jiarpakdee2020empirical}, continuous Integration prediction~\cite{zampetti2019study}, software defect developer recommendation~\cite{anvik2006should}, code reviewer recommendation~\cite{zanjani2015automatically}, etc. The interpretability of AI/ML model has naturally become the focus of these studies. According to ~\cite{miller2019explanation}, the interpretability of AI/ML model is the degree to which a human can understand the reasons behind a decision. For example, the model interpretability should reflect the relationship of feature on the outcome, the importance of each feature, the decision rule of each feature, etc~\cite{jiarpakdee2021practitioners}. The importance of model interpretability  in software engineering is obvious since without proper understanding towards the model, practitioners may not trust and adopt the model in practice~\cite{tantithamthavorn2021explainable}. 
\emph{HGRec} models the review history using hypergraph, which explicitly includes the interaction among contributors, \emph{PR}s and reviewers. 
Besides, the setting of parameters directly reflects the recommendation inclination. These characteristics of \emph{HGRec} obviously make the rationale of reviewer recommendation much easier to be understood.

\vspace{-0.1cm}
\subsection{Capability to Support Future Improvements}
Hypergraph distinguishes itself not only by its straight adaption to code review but also its architecture's flexibility and extensibility, which supplies a promotion for improvement in the future. For example, due to its architecture and search strategies,~\emph{CN} and \emph{EARec} hardly make any adjustments. 
On the contrary, \emph{HGRec} at this stage has presented the advantages to model review history using the hypergraph technique, yet it still has the capability to involve more entities and relationships, which is worthy of exploration. For example, potential reviewers belonging to the same organization may share a similar background, thus impacting the weight ($w$) calculation. In addition, the content of review comments may bring a new feature to characterize a \emph{PR}.

 To conclude,~\emph{HGRec} supports a flexibility to adjust diverse contexts and future improvements.

\subsection{Recommender Selection}

\begin{figure}[!htp]
\centering
\includegraphics[width=0.85\linewidth]{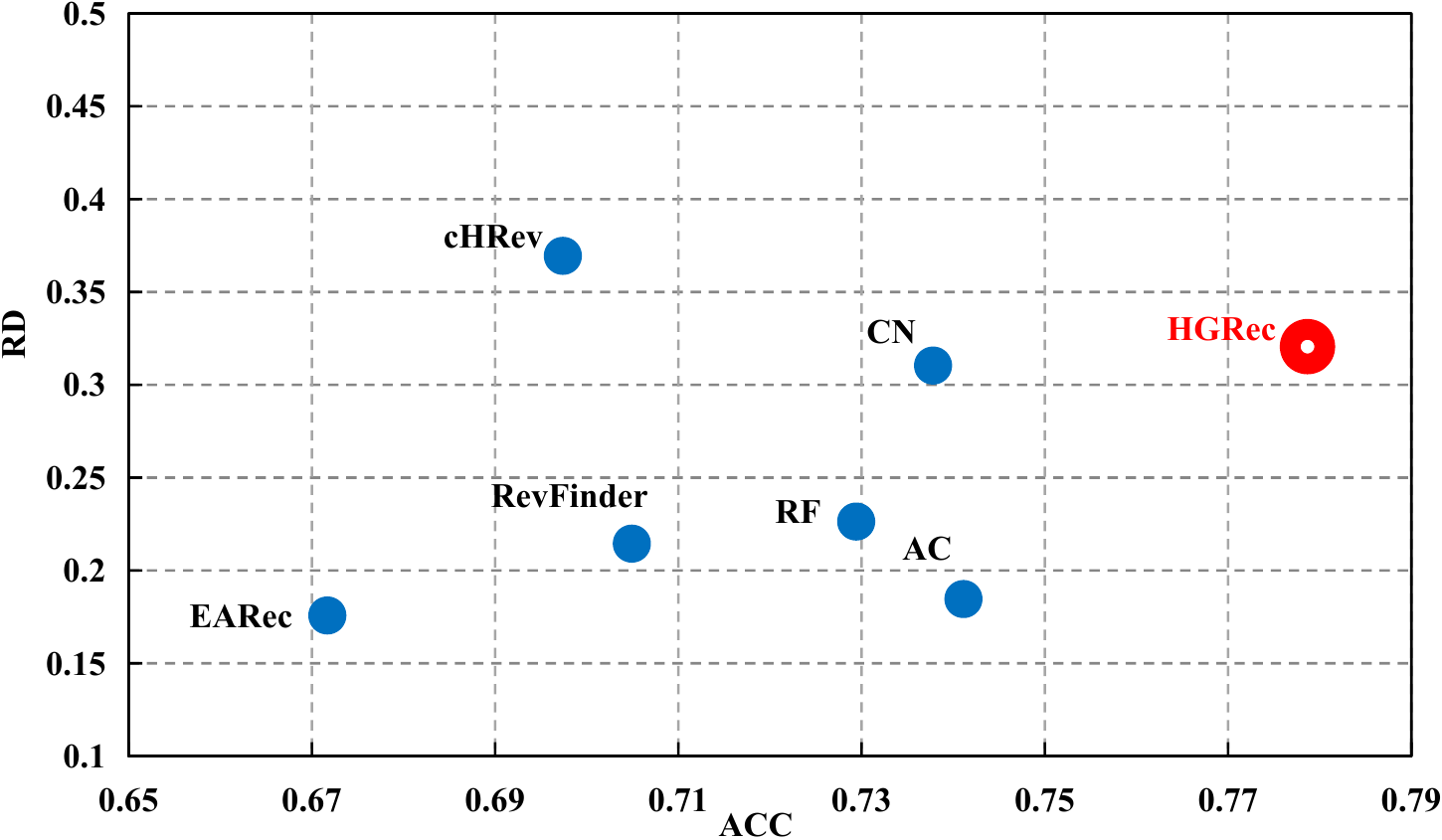}
\caption{Synthetical evaluation regarding \emph{ACC} and \emph{RD} }
\label{FIG:top3AccEnt}
\end{figure}

We use the top-3 accuracy, a common way to evaluate recommenders, combined with recommendation distribution to visually illustrate the performance of all the recommenders involved in our study. Figure~\ref{FIG:top3AccEnt}. presents the average \emph{ACC} and \emph{RD} for the 12 projects. The advantages of \emph{HGRec} are thus easy to identify, i.e. it achieves the most accurate recommendation in all recommenders and the best balanced workload in the top three accurate recommenders.

\begin{figure}[!htp]
\centering
\includegraphics[width=0.9\linewidth]{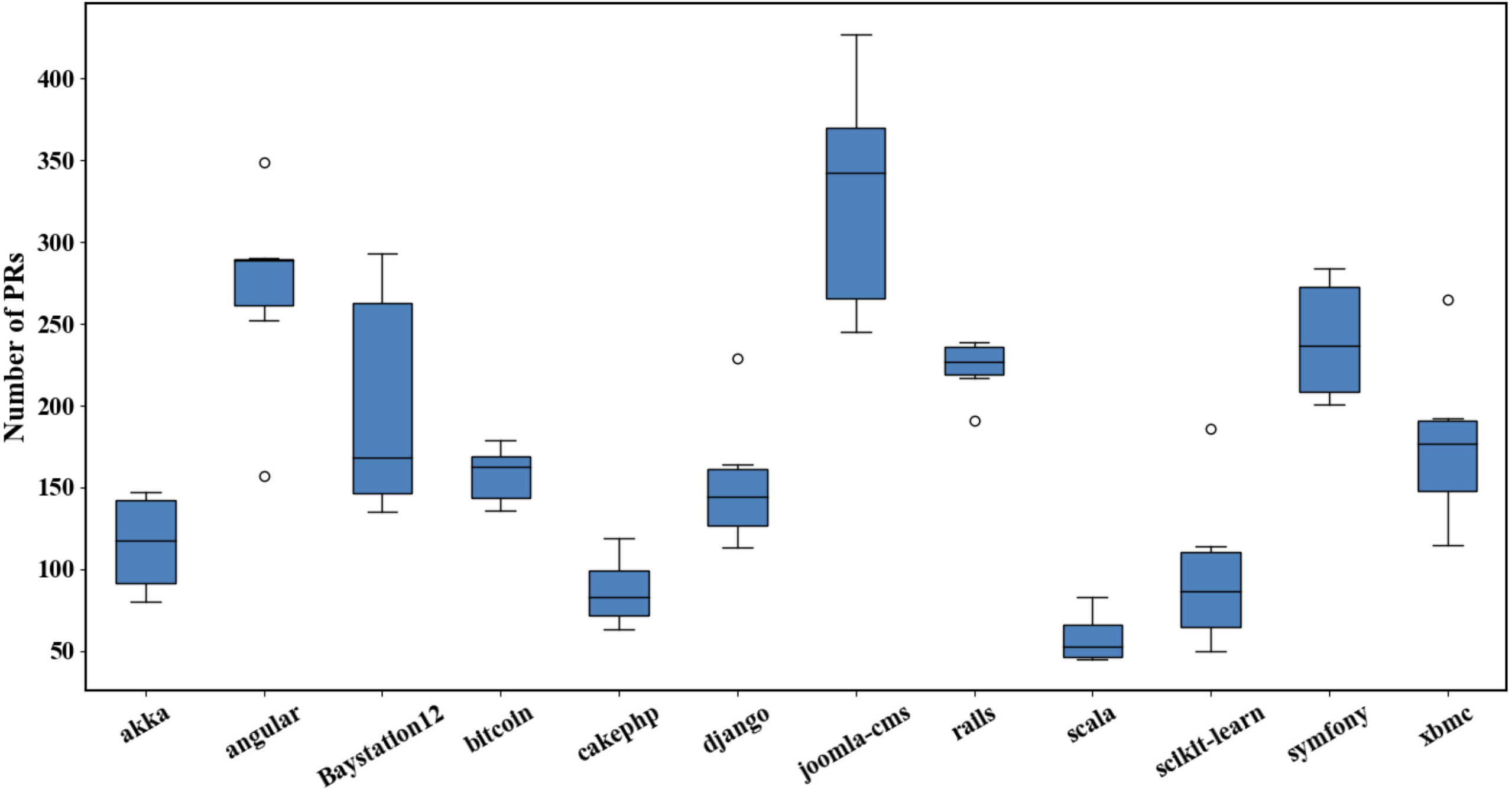}
\caption{Monthly \#\emph{PR}s of the 12 projects}
\label{FIG:projectScale}
\end{figure}

Nevertheless, although recommendation accuracy should be the primary consideration in most cases (otherwise the value of recommendation will be lost), reviewer recommendation should be applied with sufficient considerations in the application context, which involves multiple factors such as the number of potential reviewers, the number of \emph{PR}s, etc. Take project `angular' for example (as shown in Figure~\ref{fig:recommendation_distribution}), workload balance may not be an ignorable factor since the core reviewers have already undergone heavy review tasks. To present a general concept, we portray the number of \emph{PR}s per month for the 12 projects involved in our study, as depicted in Figure~\ref{FIG:projectScale}. Obviously, recommendation distribution (aka, workload balance) means more in projects such as `angular', `joomla-cms', etc., where there are normally hundreds of newly-submitted~\emph{PR}s need to be reviewed. On the contrary, in projects such as `cakephp' and `scala' where there are usually only dozens of new \emph{PR}s per month, the core two or three reviewers to handle all the \emph{PR}s may seem to be acceptable.

\subsection{Threats to Validity}
Several threats to validity are elaborated in this subsection.

\subsubsection{Construct validity}
The threats to the construct validity of this study may be related to one of the common concerns of research on reviewer recommendation, i.e., the ground-truth set of reviewers for evaluation\cite{ouni2016search}. The actual reviewers recorded in review history may not be able to guarantee ``suitable reviewers'' and further justify an appropriate recommendation. In this sense,  the recommended reviewers are only potentially suitable reviewers for a certain \emph{PR}.

\subsubsection{Internal validity}
The threats to the internal validity of this study may result from the data preparation phase. The personal organization of OSS projects is significantly loose, which brings participants' frequent turnovers and once-in-all reviews. Recommending these gone or accidental reviewers is inappropriate. In this study, we left out reviewers who had already deleted accounts or participated in less than two reviews, so did the robot users. On the other hand, the opening \emph{PR}s were also removed as they are uncertain. Another related threat is that some noise data (e.g., casual/superficial comments such as ``\emph{OK}'', ``\emph{fine}'', etc.) exists in both the training set and test set, which may not be able to guarantee a qualified reviewer recommendation. 
However, the evaluation on \emph{HGRec} and other recommenders is based on the same dataset, which may mitigate these threats to a fair degree. Meanwhile, reviewers who posted these casual/superficial comments may also have subtle relationships (e.g., certain familiarity, mutual influence, etc.). Therefore, we did not refine the dataset to remove noise data at this stage. Instead, \emph{HGRec} takes the advantage to use the possible relationships behind the casual/superficial comments and their corresponding reviewers.


\subsubsection{External validity}
We experimented with the proposed recommender on the 12 OSS projects that are retrieved from GitHub. However, the proposed recommender could suffer risks on external validity, as several studies investigated other contexts, e.g., Gerrit projects or mixed projects (both OSS and industrial projects). Therefore, the findings and conclusions are only valid in the given context. We have confidence that this study is representative because all the included projects were mentioned in the previous studies and the data is up-to-date. Besides, given the population of OSS projects from GitHub, 12 projects tested in our study may only represent a small portion. Nevertheless, the comparably consistent performance (i.e., recommendation accuracy) in our study and previous studies is able to mitigate this external threat to validity to a fair degree.

\subsubsection{Conclusion validity}
To avoid threats to conclusion validity, we followed a systematic, rigorous experiment and analysis procedure. The recommender proposed in this study has been experimented on 12 OSS projects with the history in three and a half years, including more than 87K \emph{PR}s, 680K review comments. All the dataset is clearly elaborated (e.g., the name of projects, the time range, etc.) and publicly accessible online.
This ensures a high degree of reliability that the conclusion drawn in the study is directly traceable to the raw data and hence can be replicated by other researchers.

\section{Conclusions}
With the proliferation of the pull-request development paradigm nowadays, as a key and perhaps daily practice, Modern Code Review (MCR) may impact massive software projects. While the importance of suitable reviewers has been widely recognized among OSS projects, their identification is indeed a challenge. Although many recommenders have been proposed in the past decade, their adoption is far from satisfactory~\cite{zanjani2015automatically}. Several critical issues such as low accuracy, workload congestion, incapable of extension and improvement have been raised and investigated in several related studies.

This paper proposes~\emph{HGRec}, a hypergraph based recommender to perform automatic reviewer recommendation in OSS projects. By applying hypergraph, we managed to model high-order relationships in MCR, an essential step in OSS development. A relatively extensive evaluation based on 12 OSS projects with more than 87K \emph{PR}s and 680K comments indicates that the proposed approach (i.e. ~\emph{HGRec}) outperforms the state-of-the-art recommenders in terms of accuracy. Moreover, among the top-3 accurate recommenders,~\emph{HGRec} is more likely to recommend new reviewers out of core reviewers, which may help to alleviate the workload congestion issue to some extent. Last but not least, with flexible and natural model architecture,~\emph{HGRec} can support modeling more elements (e.g., entities, attributes and relationships) in a way that more modern learning techniques or sophisticated heuristic algorithms could be incorporated into the recommender. To this end, better performance can be expected with exploration in the future.

\section*{Acknowledgments}
This work is supported by the National Natural Science Foundation of China (No.62072227, No.61802173), the National Key Research and Development Program of China (No.2019YFE0105500) jointly with the Research Council of Norway (No.309494), the Key Research and Development Program of Jiangsu Province (No.BE2021002-2), as well as the Intergovernmental Bilateral Innovation Project of Jiangsu Province (No.BZ2020017).



\balance
\bibliographystyle{ACM-Reference-Format}
\bibliography{main}

\end{document}